\documentclass[prb,preprint,superscriptaddress,preprintnumbers,amsmath,amssymb,floatfix]{revtex4}
\usepackage{graphicx}
\begin{document}
\title{Generalized kinetic equations for charge carriers in graphene}
\author{M. Auslender}
\affiliation{Ben-Gurion University of the Negev, POB 653, Beer
Sheva 84105, Israel}
\author{M. I. Katsnelson}
\affiliation{Institute for Molecules and Materials, Radboud
University of Nijmegen, NL-6525 ED Nijmegen, The Netherlands}
\date{\today}

\pacs{73.50.Bk, 72.10.Fk, 05.60.Gg, 81.05.Uw}

\begin{abstract}
A system of generalized kinetic equations for the distribution
functions of two-dimensional Dirac fermions scattered by
impurities is derived in the Born approximation with respect to
short-range impurity potential. It is proven that the conductivity
following from classical Boltzmann equation picture, where
electrons or holes have scattering amplitude reduced due
chirality, is justified except for an exponentially narrow range
of chemical potential near the conical point. When in this range,
creation of infinite number of electron-hole pairs related to
quasi-relativistic nature of electrons in graphene results in a
renormalization of minimal conductivity as compared to the
Boltzmann term and logarithmic corrections in the conductivity
similar to the Kondo effect.
\end{abstract}

\maketitle

\section*{Introduction}

Recent discovery of two-dimensional (2D) allotrope of carbon,
graphene, and experimental demonstration of its massless Dirac
energy spectrum has initiated a huge experimental and theoretical
activity in the field (for review, see Refs.\onlinecite{r1, r2,
r3}). One of the most interesting aspects of the graphene physics
from theoretical point of view is a deep and fruitful relation with
the quantum electrodynamics.
\cite{semenoff,haldane,gonzales,zitter,ktsn,falko,r4} In particular,
anomalous transport properties of 2D Dirac fermions, such as finite
conductivity of order of $e^2/h$ in the limit of zero charge carrier
concentration \cite{D1,D2,ludwig,D3,D4,shon} can be associated with
a specific quantum relativistic phenomenon known as Zitterbewegung.
\cite{zitter} The current operator of non-relativistic electron
commutes with its kinetic-energy Hamiltonian and does not commute
with the potential-energy one. Yet, it is vice versa for the Dirac
electrons that is a reason for the Zitterbewegung. The same
commutation properties hold for graphene in the case where the
potential does not cause Umklapp process. Qualitatively, an impurity
potential acting on non-relativistic electron creates random
friction-like force which causes finite conductivity. This is
expressed quantitatively in the standard theory of electronic
transport in disordered metals and semiconductors
\cite{kubo,nakano,mori,ziman} by deriving and solving the classical
Boltzmann equation. The impurity potential action on the Dirac
electron can not be described within such a simple picture. Despite
this important difference many authors exploited the classical
Boltzmann equation to analyze electron transport in graphene.
\cite{r4,shon,nomura,ando,sarma,ripples} Rigorously speaking, it is
not clear what will be the limits of its applicability in this
unusual situation. Our work presents a consequent derivation of
kinetic equations for the 2D massless Dirac fermions. Some of our
results for the static conductivity are similar to those obtained by
various quantum-field theory methods. \cite{D1,D2,ludwig,D3,D4,shon}
The approach based on the kinetic equation provides an alternative
view on the anomalous transport properties of graphene. It can be
easier generalized for more complicated situations such as strong
electric fields, hot electrons, etc. These issues are beyond the
scope of the present work. We will not consider also the effects of
Anderson localization and antilocalization
\cite{altshuler,mirlin,efetov,altland} in graphene restricting
ourselves by the case of a weak disorder in the leading-order
approximation. As we will see even in this case the problem turns
out to be very nontrivial and instructive. We will prove that for
not too small doping the standard Boltzmann equation with the
scattering amplitude specific for massless fermions does give the
leading term in the conductivity and will find corrections to it due
to the Zitterbewegung. In particular, these corrections have an
interesting temperature dependence similar to the Kondo effect.

A general idea of the approach used here is traced back to seminal
papers by Kohn and Luttinger. \cite{kohn} Starting from
Schr\"{o}dinger equation for noninteracting electrons in a random
impurity potential they consequently derived the kinetic equation
for diagonal (in momentum representation) matrix elements of the
one-electron density matrix in the cases of weak potential or small
impurity concentration. In these cases the kinetic equation turned
out to be identical with the classical Boltzmann equation. Even for
the simplest system to which it was initially applied, the Kohn and
Luttinger treatment \cite{kohn} proved not simple. For
multi-component systems one may also follow the route of
Ref.\onlinecite{kohn} and infer on existence of a closed system for
distribution functions in the momentum space - usual ones and
functions that describe inter-subsystems transitions - but
complexity of deriving such kinetic equations sharply increases.

Several established formalisms exist nowadays, which automate the
above derivation assuming existence of some kinetic equations in
principle. A partial list includes Kadanoff-Baym \cite{kadan},
Keldysh \cite{keldysh1, keldysh2}, Zubarev nonequilibrium
statistical operator (NSO) \cite{zubarev} (for the NSO method, see
also recent reviews \cite{ramos, kuzem}) and Peletminskii-Yatsenko
\cite{akhiez} methods. The Keldysh, NSO and Peletminskii-Yatsenko
methods have close rationales. Namely, existence of an asymptotic
density matrix which allows for Wick-rule decoupling of the creation
and annihilation operators product averages is assumed in these
methods. The consideration of non-equilibrium at strong interactions
benefits using the Keldysh method which is distinguished for highly
developed diagram technique. At weak interactions, however, when the
Born approximation is applicable the simplest approach in our
opinion is with the NSO and Peletminskii-Yatsenko methods. This is
because in the Born approximation, closed equations for the averages
of gross variables, generalized kinetic equations (GKE), which
describe non-equilibrium of interest (provided that such variables
are declared in advance) were derived within these frameworks in
late 60's once for all. \cite{zubarev, akhiez, ramos, kuzem}

In this paper we obtain and asymptotically solve GKE for spatially
homogeneous graphene in order to calculate the linear-response
conductivity. The main difference with the canonical case
\cite{kohn} is that for graphene the diagonal in the momentum
representation average density matrix is still two by two matrix in
the pseudospin space, its off-diagonal elements describing the
Zitterbewegung. This makes the GKE structure, on the whole,
essentially different from that of the classical kinetic equation.
The structure of the paper is the following. In section I we present
original expressions for the Hamiltonian, current and coordinate
operators. In section II we specify the gross variables appropriate
for the kinetics in spatially homogeneous case, which in fact are
all the density matrix elements, and on their base concretize GKE
regarding the interaction with arbitrary static impurities as a
perturbation. In section III, assuming presence of a thermostat, we
consider linear response regime (the case of small electric field)
and express the linear static conductivity via two unknown functions
of the one-electron energy. These functions, together with a
subsidiary function of energy, satisfy a coupled system of linear
integral equations resulting from linearizing GKE in electric field
strength. In section IV, we solve the linearized system and
calculate the conductivity within an ultraviolet cut off Dirac-delta
impurity potential asymptotically in a controllable small parameter,
using methods of solving singular integral equations. In section V
we discuss the results obtained.

\section{Preliminaries}

We proceed with the Hamiltonian of two-dimensional massless Dirac
fermions describing charge carriers in graphene if one neglects the
Umklapp processes between valleys $K$ and $K'$
\begin{equation}
\mathcal{H}_0=v\sum_{\mathbf{p}}\Psi_{\mathbf{p}}^{\dag}\left(
\boldsymbol{\tau }\cdot\mathbf{p}\right)  \Psi_{\mathbf{p}}
\label{DiracHam}
\end{equation}
where $\mathbf{p}$ is the momentum vector, $v$ is the velocity,
\begin{equation}
\Psi_{\mathbf{p}}=\left(
\begin{array}
[c]{c}
\psi_{\mathbf{p}1}\\
\psi_{\mathbf{p}2}
\end{array}
\right)  ,\,\Psi_{\mathbf{p}}^{\dag}=\left(
\begin{array}
[c]{cc} \psi_{\mathbf{p}1}^{\dag} & \psi_{\mathbf{p}2}^{\dag}
\end{array} \right)  \label{Bare-electr}
\end{equation}
are two-component pseudospinor operators, 1,2 labelling the
sublattices, and
\begin{equation}
\boldsymbol{\tau}=\left(  \tau_{x},\tau_{y}\right)  \,,\,\tau_{x}=\left(
\begin{array}
[c]{cc}
0 & 1\\
1 & 0
\end{array}
\right)  ,\tau_{y}=\left(
\begin{array}
[c]{cc}%
0 & -i\\
i & 0
\end{array}
\right)  \label{Pseudospin}
\end{equation}
are the Pauli matrices in the pseudospin space. We will neglect
here real spin and valley indices. The Hamiltonian
(\ref{DiracHam}) can be diagonalized using the unitary
transformation matrix~\cite{shon}
\begin{equation}
U_{\mathbf{p}}=\frac{1}{\sqrt{2}}\left(
\begin{array}
[c]{cc}%
1 & 1\\
e^{i\phi_{\mathbf{p}}} & -e^{i\phi_{\mathbf{p}}}%
\end{array}
\right)
\end{equation}
where $\phi_{\mathbf{p}}$ is the polar angle of the vector
$\mathbf{p}$.
Hence the new electron operators given by%
\begin{equation}
\Xi_{\mathbf{p}}=U_{\mathbf{p}}^{\dag}\Psi_{\mathbf{p}}
=\frac{1}{\sqrt{2}}\left(
\begin{array}
[c]{c}%
\psi_{\mathbf{p}1}+e^{-i\phi_{\mathbf{p}}}\psi_{\mathbf{p}2}\\
\psi_{\mathbf{p}1}-e^{-i\phi_{\mathbf{p}}}\psi_{\mathbf{p}2}%
\end{array}
\right)  =\left(
\begin{array}
[c]{c}%
\xi_{\mathbf{p}1}\\
\xi_{\mathbf{p}2}
\end{array}
\right)
\end{equation}
and%
\begin{align}
\Xi_{\mathbf{p}}^{\dag} &
=\Psi_{\mathbf{p}}^{\dag}U_{\mathbf{p}}=\left(
\begin{array}
[c]{cc}%
\xi_{\mathbf{p}1}^{\dag} & \xi_{\mathbf{p}2}^{\dag}%
\end{array}
\right)
\end{align}
are the annihilation and creation operators of the conduction and valence band
electrons. Thus we have%
\begin{equation}
\Psi_{\mathbf{p}}=U_{\mathbf{p}}\Xi_{\mathbf{p}},\,\Psi_{\mathbf{p}}^{\dag
}=\Xi_{\mathbf{p}}^{\dag}U_{\mathbf{p}}^{\dag}%
\end{equation}
and
\begin{equation}
\mathcal{H}_0=v\sum_{\mathbf{p},s=\pm1}sp\xi_{\mathbf{p}s}^{\dag}\xi
_{\mathbf{p}s}.
\end{equation}

In what follows we will consider the simplest case where electrons
experience action of a scalar potential $V\left (\mathbf{r}
\right)$ presents. The interaction Hamiltonian in this case is
given by
\begin{equation}
\mathcal{H}_{\text{int}}  = S^{-1}\sum_{\mathbf{pp}^{\prime}}V\left(  \mathbf{p}%
-\mathbf{p}^{\prime}\right)
\Psi_{\mathbf{p}}^{\dag}\Psi_{\mathbf{p}^{\prime }} =
S^{-1}\sum_{\mathbf{pp}^{\prime}}\Xi_{\mathbf{p}
}^{\dag}\widehat{V}_{\mathbf{pp}^{\prime}}\Xi_{\mathbf{p}^{\prime}}%
\end{equation}
where $S$ is the graphene layer surface area, $V\left (\mathbf{q}
\right)$ is the Fourier transform of $V\left (\mathbf{r} \right)$
and
\begin{equation}
\widehat{V}_{\mathbf{pp}^{\prime}}=\frac{1}{2}V\left(  \mathbf{p}
-\mathbf{p}^{\prime}\right)  \left(
\begin{array}
[c]{cc}%
1+e^{-i\left(  \phi_{\mathbf{p}}-\phi_{\mathbf{p}^{\prime}}\right)  } &
1-e^{-i\left(  \phi_{\mathbf{p}}-\phi_{\mathbf{p}^{\prime}}\right)  }\\
1-e^{-i\left(  \phi_{\mathbf{p}}-\phi_{\mathbf{p}^{\prime}}\right)  } &
1+e^{-i\left(  \phi_{\mathbf{p}}-\phi_{\mathbf{p}^{\prime}}\right)  }%
\end{array} \right).
\end{equation}

The current density operator in the new variables reads
\begin{equation}
\mathbf{J}=ev\sum_{\mathbf{p}}\Psi_{\mathbf{p}}^{\dag}\boldsymbol{\tau}%
\Psi_{\mathbf{p}}=\sum_{\mathbf{p}}\Xi_{\mathbf{p}}^{\dag}\mathbf{j}%
_{\mathbf{p}}\Xi_{\mathbf{p}},
\end{equation}
where%
\begin{equation}
\mathbf{j}_{\mathbf{p}}=evU_{\mathbf{p}}^{\dag}\boldsymbol{\tau}U_{\mathbf{p}%
}=\left(  j_{\mathbf{p}x},j_{\mathbf{p}y}\right)  ,
\end{equation}
and%
\begin{eqnarray}
j_{\mathbf{p}x}  & = ev\left(
\begin{array}
[c]{cc}%
\cos\phi_{\mathbf{p}} & -i\sin\phi_{\mathbf{p}}\\
i\sin\phi_{\mathbf{p}} & -\cos\phi_{\mathbf{p}}%
\end{array}
\right),\hspace{0.25 cm} j_{\mathbf{p}y} = ev\left(
\begin{array}
[c]{cc}
\sin\phi_{\mathbf{p}} & i\cos\phi_{\mathbf{p}}\\
-i\cos\phi_{\mathbf{p}} & -\sin\phi_{\mathbf{p}}%
\end{array}
\right)  .
\end{eqnarray}
Off-diagonal elements of the current operator correspond to the
Zitterbewegung processes. \cite{zitter} For the $x$ and
$y$-components of the current, we further obtain
\begin{align}
J_{x}  &  = ev\sum_{\mathbf{p}}\Xi_{\mathbf{p}}^{\dag}\left(
\begin{array}
[c]{cc}
\cos\phi_{\mathbf{p}} & -i\sin\phi_{\mathbf{p}} \nonumber \\
i\sin\phi_{\mathbf{p}} & -\cos\phi_{\mathbf{p}}%
\end{array}
\right)  \Xi_{\mathbf{p}} \nonumber \\
&  =ev\sum_{\mathbf{p}}\left[  \cos\phi_{\mathbf{p}}\left(  \xi_{\mathbf{p,}%
1}^{\dag}\xi_{\mathbf{p,}1}-\xi_{\mathbf{p,}-1}^{\dag}\xi_{\mathbf{p,}%
-1}\right)  -i\sin\phi_{\mathbf{p}}\left(  \xi_{\mathbf{p,}1}^{\dag}%
\xi_{\mathbf{p,}-1}-\xi_{\mathbf{p,}-1}^{\dag}\xi_{\mathbf{p,}1}\right)
\right],
\end{align}
and
\begin{align}
J_{y}  &  = ev\sum_{\mathbf{p}}\Xi_{\mathbf{p}}^{\dag}\left(
\begin{array}
[c]{cc}
\sin\phi_{\mathbf{p}} & i\cos\phi_{\mathbf{p}} \nonumber \\
-i\cos\phi_{\mathbf{p}} & -\sin\phi_{\mathbf{p}}%
\end{array}
\right)  \Xi_{\mathbf{p}} \nonumber \\
&  =ev\sum_{\mathbf{p}}\left[  \sin\phi_{\mathbf{p}}\left(  \xi_{\mathbf{p,}%
1}^{\dag}\xi_{\mathbf{p,}1}-\xi_{\mathbf{p,}-1}^{\dag}\xi_{\mathbf{p,}%
-1}\right)  +i\cos\phi_{\mathbf{p}}\left(  \xi_{\mathbf{p,}1}^{\dag}%
\xi_{\mathbf{p,}-1}-\xi_{\mathbf{p,}-1}^{\dag}\xi_{\mathbf{p,}1}\right)
\right]. \label{jx}
\end{align}

At last, the electron coordinate operator which is necessary to
derive the field term in the kinetic equation reads
\begin{equation}
\mathbf{R}=i\sum_{\mathbf{p}}\Psi_{\mathbf{p}}^{\dag}\boldsymbol{\nabla}%
\Psi_{\mathbf{p}},
\end{equation}
where $\nabla$ is the gradient operator with respect to the
momentum $\mathbf{p}$. Using the above unitary transformation, we
get
\begin{align}
\mathbf{R}  &  =i\sum_{\mathbf{p}}\Xi_{\mathbf{p}}^{\dag}U_{\mathbf{p}}^{\dag
}\boldsymbol{\nabla}\left(  U_{\mathbf{p}}\Xi_{\mathbf{p}}\right)
=i\sum_{\mathbf{p}}\Xi_{\mathbf{p}}^{\dag}\boldsymbol{\nabla}\Xi_{\mathbf{p}%
}+i\sum_{\mathbf{p}}\Xi_{\mathbf{p}}^{\dag}U_{\mathbf{p}}^{\dag}\left
[ \boldsymbol{\nabla}\frac{1}{\sqrt{2}}\left(
\begin{array}
[c]{cc}
1 & 1\\
e^{i\phi_{\mathbf{p}}} & -e^{i\phi_{\mathbf{p}}}
\end{array}
\right)  U_{\mathbf{p}}\right ]  \Xi_{\mathbf{p}} \nonumber \\
&  =\sum_{\mathbf{p}}\left[  i\left(  \xi_{\mathbf{p}1}^{\dag}%
\boldsymbol{\nabla}\xi_{\mathbf{p}1}+\xi_{\mathbf{p},-1}^{\dag}%
\boldsymbol{\nabla}\xi_{\mathbf{p},-1}\right)  -\frac{1}{2}\left(
\xi_{\mathbf{p}1}^{\dag}\xi_{\mathbf{p}1}+\xi_{\mathbf{p},-1}^{\dag}%
\xi_{\mathbf{p},-1}\right)  \boldsymbol{\nabla}\phi_{\mathbf{p}}\right] \nonumber \\
&  +\frac{1}{2}\sum_{\mathbf{p}}\boldsymbol{\nabla}\phi_{\mathbf{p}}\left(
\xi_{\mathbf{p}1}^{\dag}\xi_{\mathbf{p},-1}+\xi_{\mathbf{p},-1}^{\dag}%
\xi_{\mathbf{p}1}\right).
\end{align}
To simplify $\mathbf{R}$ let us perform additional gauge
transformation
$\xi_{\mathbf{p}s}\rightarrow e^{-i\frac{1}{2}\phi_{\mathbf{p}}}%
\xi_{\mathbf{p}s}$ that retains the Hamiltonian $\mathcal{H}_0$
unchanged but renormalizes the coordinate and current density
operators to take the form
\begin{equation}
\mathbf{R}\rightarrow\sum_{\mathbf{p},m=\pm1}\left(  i\xi_{\mathbf{p}m}^{\dag
}\boldsymbol{\nabla}\xi_{\mathbf{p}m}+\frac{1}{2}\xi_{\mathbf{p}m}^{\dag}%
\xi_{\mathbf{p},-m}\boldsymbol{\nabla}\phi_{\mathbf{p}}\right)  =\mathbf{R}%
_{\text{intra}}+\mathbf{R}_{\text{inter}}%
\end{equation}
and
\begin{equation}
\mathbf{J}\rightarrow ev\sum_{\mathbf{p},m=\pm1}m\left(  \frac{\mathbf{p}}%
{p}\xi_{\mathbf{p}m}^{\dag}\xi_{\mathbf{p}m}+i\xi_{\mathbf{p}m}^{\dag}%
\xi_{\mathbf{p},-m}p\boldsymbol{\nabla}\phi_{\mathbf{p}}\right)
=\mathbf{J}_{\text{intra}}+\mathbf{J}_{\text{inter}},
\end{equation}
respectively. Here we have separated explicitly intraband
(electron-electron and hole-hole) and interband (electron-hole)
contributions. Note that
\begin{equation}
\frac{d}{dt}e\mathbf{R}=\mathbf{J,}%
\end{equation}
as it should be. The interaction matrix elements are thus
transformed to
\begin{equation}
\widehat{V}_{\mathbf{pp}^{\prime}}\rightarrow V\left(  \mathbf{p}%
-\mathbf{p}^{\prime}\right)  \left(
\begin{array}
[c]{cc}%
\cos\frac{\phi_{\mathbf{p}}-\phi_{\mathbf{p}^{\prime}}}{2} & i\sin\frac
{\phi_{\mathbf{p}}-\phi_{\mathbf{p}^{\prime}}}{2}\\
i\sin\frac{\phi_{\mathbf{p}}-\phi_{\mathbf{p}^{\prime}}}{2} & \cos\frac
{\phi_{\mathbf{p}}-\phi_{\mathbf{p}^{\prime}}}{2}%
\end{array}
\right). \label{Vroof}
\end{equation}

\section{THE BORN-APPROXIMATION KINETIC EQUATIONS}

\subsection{General outline}

The basic idea of the methods of Refs.\onlinecite{zubarev, akhiez}
is a concept of so called ``coarse-grained'' dynamics. To apply the
formalism we are, as noted in Introduction, to suggest the gross
variables $P$, averages of which $\left\langle P \right\rangle $ at
the kinetic stage of the evolution are believed to satisfy GKE. It
was proven by Kohn and Luttinger \cite{kohn} that, if $V\left(
\mathbf{r} \right)$ is due to random impurities, the diagonal
elements of the one-electron density matrix in the momentum
representation averaged over weakly perturbed non-equilibrium
ensemble are self averaging over the impurity configurations and do
obey such a reduced description, at least for weak enough potential
or small impurity concentration. Our problem is formally different
from standard one only in existence of the interband operators.
Therefore we choose the following gross variables
\begin{equation}
P_{\mathbf{p}}=\left(
\begin{array}
[c]{c}%
\xi_{\mathbf{p}1}^{\dag}\xi_{\mathbf{p}1}\\
\xi_{\mathbf{p},-1}^{\dag}\xi_{\mathbf{p},-1}\\
\xi_{\mathbf{p}1}^{\dag}\xi_{\mathbf{p},-1}\\
\xi_{\mathbf{p},-1}^{\dag}\xi_{\mathbf{p}1}%
\end{array} \right),
\end{equation}
the components of this vector being the second-quantization form
of the above matrix elements. The corresponding
``quasi-equilibrium'' or ``coarse-grained'' statistical operator
(QSO) \cite{zubarev, akhiez} is given by
\begin{equation}
\rho_{\text
q}=e^{-\Phi-\sum\limits_{\mathbf{p}}F_{\mathbf{p}}^{\dag}P_{\mathbf{p}}
}=e^{-\Phi-\sum\limits_{\mathbf{p}}\left(
F_{\mathbf{p}1}\xi_{\mathbf{p}
1}^{\dag}\xi_{\mathbf{p}1}+F_{\mathbf{p}2}\xi_{\mathbf{p},-1}^{\dag}
\xi_{\mathbf{p},-1}+F_{\mathbf{p}}3\xi_{\mathbf{p}1}^{\dag}\xi_{\mathbf{p}%
,-1}+F_{\mathbf{p}3}^{\ast}\xi_{\mathbf{p},-1}^{\dag}\xi_{\mathbf{p}1}\right)
}  \label{quasi},
\end{equation}
where
\begin{equation}
\Phi=\ln\mbox{Tr}\left[  e^{-\sum\limits_{\mathbf{p}}\left(  F_{\mathbf{p}%
1}\xi_{\mathbf{p}1}^{\dag}\xi_{\mathbf{p}1}+F_{\mathbf{p}2}\xi_{\mathbf{p}%
,-1}^{\dag}\xi_{\mathbf{p},-1}+F_{\mathbf{p}3}\xi_{\mathbf{p}1}^{\dag}%
\xi_{\mathbf{p},-1}+F_{\mathbf{p}3}^{\ast}\xi_{\mathbf{p},-1}^{\dag}%
\xi_{\mathbf{p}1}\right)  }\right]
\end{equation}
is the generalized Masseu-Plank function. As at the equilibrium,
$F_{\mathbf{p} }^{\dag}$ are parameters conjugated to
$P_{\mathbf{p}}$ in the sense that \cite{zubarev, ramos, kuzem,
akhiez}
\begin{equation}
\left\langle P_{\mathbf{p}}\right\rangle _{\text q} =
-\frac{\delta \Phi}{\delta F_{\mathbf{p}}^{\dag}}.
\end{equation}
This QSO is second-quantization representation form of the general
density matrix in the which allows for Wick rules. Following from it
explicit connection between $\left\langle
P_{\mathbf{p}}\right\rangle $ and $F_{\mathbf{p}}$, however, bears
no new information.

To obtain GKE one uses the closure condition $\left\langle
P_{\mathbf{p}}\right\rangle _{\text q} = \left\langle
P_{\mathbf{p}}\right\rangle $ assumed \cite{zubarev, akhiez}  only
for the gross variables, which results in
\begin{equation}
\frac{\partial}{\partial t}\left\langle P\right\rangle _{\text
q}=i\left\langle \left[ \mathcal{H}_{0}+\mathcal{H}_{\text
{int}},P\right]  \right\rangle, \label{NSOclosure}
\end{equation}
where the averaging in the right-hand sides is performed over NSO
obtained from QSO via an explicit formal prescription
\cite{zubarev, akhiez}. In our case this averaging is also to
incorporate one over the impurity configurations. Note that in all
known cases with weak interaction the operators $P_{\mathbf{p}}$
obey closed microscopic dynamics with the unperturbed Hamiltonian
\begin{equation}
\left [\mathcal{H}_{0}, P_{\mathbf{p}} \right] = \sum_{\mathbf{q}}
\omega_{\mathbf{p}\mathbf{q}}P_{\mathbf{q}}, \label{cldyn}
\end{equation}
where $\omega_{\mathbf{p}\mathbf{q}}$ is a known matrix. It can be
shown that in our case Eq. (\ref{cldyn}) sustains even if
$\mathcal{H}_{0}$ includes the interaction with an electric field
$\mathbf{E}$ along the $x$-axis
\begin{equation}
\mathcal{H}_{\text{ef}}=-e\mathbf{E}\cdot\mathbf{R} =-eE\sum_{\mathbf{p},m=\pm1}\left(  i\xi_{\mathbf{p}%
m}^{\dag}\boldsymbol{\nabla}\xi_{\mathbf{p}m}-\frac{\sin\phi_{\mathbf{p}}}%
{2p}\xi_{\mathbf{p}m}^{\dag}\xi_{\mathbf{p},-m}\right).
\end{equation}
At that occurrence, the matrix $\omega_{\mathbf{p}\mathbf{q}}$
contains linear in $E$ off-diagonal elements, some of which
involve the gradients of the momentum-conservation delta function.

In the second-order approximation with respect to
$\mathcal{H}_{\text {int}}$ Eq. (\ref{NSOclosure}) can be
transformed \cite{zubarev, kuzem, akhiez} to GKE, which have the
following form common to all applications
\begin{equation}
\frac{\partial\left\langle P_{\mathbf{p}}\right\rangle }{\partial
t} = i\sum\limits_{\mathbf{q}}\omega_{\mathbf{pq} }\left\langle
P_{\mathbf{q}}\right\rangle + \mathcal{J}^{(1)}_{\mathbf{p}} +
\mathcal{J}^{(2)}_{\mathbf{p}}, \label{GK}
\end{equation}
where the generalized collision integrals of the first and second
orders are given by \cite{zubarev, kuzem, akhiez}
\begin{equation}
\mathcal{J}^{(1)}_{\mathbf{p}} = i \left\langle \left [
\mathcal{H_{\text {int}}}, P_{\mathbf{p}} \right
]\right\rangle_{\text q} \label{J1}
\end{equation}
and
\begin{equation}
\mathcal{J}_{\mathbf{p}}^{\left( 2\right) }=
\lim_{\varepsilon\rightarrow+0}\int_{-\infty}^{0}e^{\varepsilon
t}dt\left\langle \left[ \mathcal{H} _{\text{int}}\left( t\right),
\left[ P_{\mathbf{p}},\mathcal{H}_{\text{int}}\right] +
i\sum_{\mathbf{l}}\frac {\delta\mathcal{J}_{\mathbf{p}}^{\left(
1\right) }}{\delta\left\langle P_{\mathbf{l}}\right\rangle
}P_{\mathbf{l}}\right] \right\rangle _{\text{q}}, \label{J2}
\end{equation}
respectively. Here the dependence of the interaction Hamiltonian on
time $t$ is according to Heisenberg picture with $\mathcal{H}_{0}$
which, in addition to the kinetic energy, may include
$\mathcal{H}_{\text{ef}}$. The term in Eq. (\ref{J2}), which
involves $\mathcal{J}_{\mathbf{p}}^{\left( 1\right) }$, leads to
cancelling possible contributions that diverge in the thermodynamic
limit $S \rightarrow \infty$ out of
$\mathcal{J}_{\mathbf{p}}^{\left( 2\right) }$. This is fair analog
(in the Born approximation) to ``connected-diagrams'' statement in
the diagram techniques.\cite{kadan, keldysh1, keldysh2}
\subsection{Average current density}
Using Eq.(\ref{jx}) the average current density can be expressed
via the basic averages $\left\langle P_{\mathbf{p}}\right\rangle$
as follows
\begin{equation}
j_{x}=\frac{\left\langle J_{x}\right\rangle }{S}=\frac{ev}{\left(
2\pi\right)  ^{2}}\int\left[  \left(  f_{\mathbf{p}1}-f_{\mathbf{p}%
,-1}\right)  \cos\phi_{\mathbf{p}}+2\operatorname{Im}\left(
g_{\mathbf{p} 1}\right)  \sin\phi_{\mathbf{p}}\right]  d^{2}p,
\label{avcurrent}
\end{equation}
where, by definition,
\begin{equation}
f_{\mathbf{p}s}=\left\langle \xi_{\mathbf{p}s}^{\dag}\xi_{\mathbf{p}%
s}\right\rangle ,\,g_{\mathbf{p}s}=\left\langle \xi_{\mathbf{p}s}^{\dag}%
\xi_{\mathbf{p},-s}\right\rangle =g_{\mathbf{p},-s}^{\ast}.
\label{fg}
\end{equation}
Let us now introduce the electron and hole distribution
functions
\begin{equation}
n_{\mathbf{p}}=f_{\mathbf{p}1},\;p_{\mathbf{p}}=1-f_{\mathbf{p,-}1},
\end{equation}
which are, of course, real, and ``anomalous'' distribution
function $g_{\mathbf{p}}=g_{\mathbf{p}1}$, which is complex in
general. In the terms of these functions Eq. (\ref {avcurrent}) is
written as follows
\begin{align}
j_{x}  &  = \frac{ev}{\left(  2\pi\right)  ^{2}}\int\left[  \left(
n_{\mathbf{q}}+p_{\mathbf{q}}\right)  \cos\phi_{\mathbf{q}}+2\operatorname{Im}%
\left(  g_{\mathbf{q}}\right)  \sin\phi_{\mathbf{q}}\right]  d^{2}q \nonumber \\
&  \equiv\frac{ev}{\left(  2\pi\right)  ^{2}}\int\left[  N_{\mathbf{q}}%
\cos\phi_{\mathbf{q}}+2\operatorname{Im}\left(
g_{\mathbf{q}}\right) \sin\phi_{\mathbf{q}}\right]  d^{2}q.
\label{jxxx}
\end{align}
\subsection{Derivation details}

Let us now specify Eqs. (\ref{GK}) - (\ref{J2}) for graphene. To
this end, consider all the prerequisites of the calculations
required. Using Eq. (\ref{Vroof}), we find
\begin{align}
\mathcal{H}_{\text{int}} & =
S^{-1}\sum_{\mathbf{ll}^{\prime}}V\left(
\mathbf{l}-\mathbf{l}^{\prime}\right)\Xi_{\mathbf{l}}^{\dag}
\left(
\begin{array}
[c]{cc}%
\cos\frac{\phi_{\mathbf{l}}-\phi_{\mathbf{l}^{\prime}}}{2} & i\sin\frac
{\phi_{\mathbf{l}}-\phi_{\mathbf{l}^{\prime}}}{2}\\
i\sin\frac{\phi_{\mathbf{l}}-\phi_{\mathbf{l}^{\prime}}}{2} & \cos\frac
{\phi_{\mathbf{l}}-\phi_{\mathbf{l}^{\prime}}}{2}%
\end{array}
\right) \Xi_{\mathbf{l}^{\prime}}\nonumber \\
&  = S^{-1}\sum_{\mathbf{ll}^{\prime},m=\pm1}V\left(
\mathbf{l}-\mathbf{l}^{\prime
}\right)  \left(  \cos\frac{\phi_{\mathbf{l}}-\phi_{\mathbf{l}^{\prime}}}%
{2}\xi_{\mathbf{l,}m}^{\dag}\xi_{\mathbf{l}^{\prime}\mathbf{,}m}++i\sin
\frac{\phi_{\mathbf{l}}-\phi_{\mathbf{l}^{\prime}}}{2}\xi_{\mathbf{l,}m}%
^{\dag}\xi_{\mathbf{l}^{\prime}\mathbf{,}-m}\right)
\end{align}
and so
\begin{align}
& \frak{L} = \int_{-\infty}^{0}e^{\varepsilon
t}\mathcal{H}_{\text{int}}\left( t\right) dt \nonumber \\
& = S^{-1}\sum_{\mathbf{ll}^{\prime},m=\pm1}V\left(
\mathbf{l-l}^{\prime}\right)
\left[  \frac{\cos\frac{\phi_{\mathbf{l}}-\phi_{\mathbf{l}^{\prime}}}{2}%
}{\varepsilon+im\left(  \epsilon_{l}-\epsilon_{l^{\prime}}\right)  }%
\xi_{\mathbf{l}m}^{\dag}\xi_{\mathbf{l}^{\prime}m}+\frac{i\sin\frac
{\phi_{\mathbf{l}}-\phi_{\mathbf{l}\prime}}{2}}{\varepsilon+im\left(
\epsilon_{l}+\epsilon_{l^{\prime}}\right)  }\xi_{\mathbf{l}m}^{\dag}%
\xi_{\mathbf{l}^{\prime},-m}\right]. \label{GotL}
\end{align}
Consider further the commutators of the gross-variable operators
with $\mathcal{H}_{0}$. By a straightforward calculation we obtain
\begin{align*}
\left [\mathcal{H}_{0}, \xi_{\mathbf{p}s}^{\dag}\xi_{\mathbf{p}s}
\right ] & =\left[
\sum_{\mathbf{q},n=\pm1}\left(  nvp\xi_{\mathbf{q}n}^{\dag}\xi_{\mathbf{q}%
n}-ieE\xi_{\mathbf{q}n}^{\dag}\frac{\partial\xi_{\mathbf{q}n}}{\partial q_{x}%
}+\frac{eE}{2q}\sin\phi_{\mathbf{q}}\xi_{\mathbf{q}n}^{\dag}\xi_{\mathbf{q}%
,-n}\right)  ,\xi_{\mathbf{p}s}^{\dag}\xi_{\mathbf{p}s}\right] \\
&  =ieE\frac{\partial}{\partial p_{x}}\left(  \xi_{\mathbf{p}s}^{\dag}%
\xi_{\mathbf{p}s}\right)  -\frac{eE}{2p}\left(  \xi_{\mathbf{p}s}^{\dag}%
\xi_{\mathbf{p},-s}-\xi_{\mathbf{p},-s}^{\dag}\xi_{\mathbf{p}s}\right)
\sin\phi_{\mathbf{p}}
\end{align*}
and
\begin{align*}
\left[
\mathcal{H}_{0},\xi_{\mathbf{p}s}^{\dag}\xi_{\mathbf{p},-s}\right
] &  =\left[
\sum_{\mathbf{q},n=\pm1}\left(  nvp\xi_{\mathbf{q}n}^{\dag}\xi_{\mathbf{q}%
n}-ieE\xi_{\mathbf{q}n}^{\dag}\frac{\partial\xi_{\mathbf{q}n}}{\partial q_{x}%
}+\frac{eE}{2q}\sin\phi_{\mathbf{q}}\xi_{\mathbf{q}n}^{\dag}\xi_{\mathbf{q}%
,-n}\right)  ,\xi_{\mathbf{p}s}^{\dag}\xi_{\mathbf{p,-}s}\right] \\
&  =2svp\xi_{\mathbf{p}s}^{\dag}\xi_{\mathbf{p,-}s}+ieE\frac{\partial
}{\partial p_{x}}\left(  \xi_{\mathbf{p}s}^{\dag}\xi_{\mathbf{p},-s}\right)
-\frac{eE}{2p}\left(  \xi_{\mathbf{p}s}^{\dag}\xi_{\mathbf{p},s}%
-\xi_{\mathbf{p,-}s}^{\dag}\xi_{\mathbf{p,-}s}\right)  \sin\phi_{\mathbf{p}}. %
\end{align*}
Hence the ``precession'' terms in the right-hand side of GKE, see
Eq. (\ref{GK}), are
\begin{align}
i\sum_{\mathbf{q}}\omega_{\mathbf{p}s,\mathbf{q}s}\left\langle
\xi_{\mathbf{p}s}^{\dag}\xi_{\mathbf{q}
s}\right\rangle _{q}  &  =-eE\frac{\partial f_{\mathbf{p}s}}{\partial p_{x}%
}+\frac{eE\sin\phi_{\mathbf{p}}}{p}\operatorname{Im}g_{\mathbf{p}s}, \nonumber \\
i\sum_{\mathbf{q}}\omega_{\mathbf{p}s,\mathbf{q}-s}\left\langle
\xi_{\mathbf{p}s}^{\dag}\xi_{\mathbf{q} s}\right\rangle _{q}  & =
2isvp\,g_{\mathbf{p}s}-eE\frac{\partial
\,g_{\mathbf{p}s}}{\partial
p_{x}}-i\frac{eE\sin\phi_{\mathbf{p}}}{2p} \left( f_{\mathbf{p}
s}-f_{\mathbf{p,-}s}\right)  \label{field},
\end{align}
where we have used the notations introduced in Eq. (\ref{fg}).

To calculate collision integrals for the ``normal'' distribution
functions $f_{\mathbf{p}s}$ and the anomalous ones
$g_{\mathbf{p}s}$ we are to perform the commutation twice - first
time to commute the gross-variables operators with
$\mathcal{H}_{\text{int}}$ to obtain
$\mathcal{J}^{(1)}_{\mathbf{p}}$ and the second to commute the
result of the first commutation with $\frak{L}$ to obtain
$\mathcal{J}^{(2)}_{\mathbf{p}}$. Following this route we get
\begin{align}
\left[  \xi_{\mathbf{p}s}^{\dag}\xi_{\mathbf{p}s},\mathcal{H}_{\text{int}%
}\right]   &  = S^{-1}\sum_{\mathbf{qq}^{\prime}}V\left(  \mathbf{q}-\mathbf{q}%
^{\prime}\right)  \left[
\cos\frac{\phi_{\mathbf{q}}-\phi_{\mathbf{q} ^{\prime}}}{2}\left(
\delta_{\mathbf{p,q}}-\delta_{\mathbf{p,q}^{\prime}
}\right)  \xi_{\mathbf{q}s}^{\dag}\xi_{\mathbf{q}^{\prime}s}\right. \nonumber \\
&  +\left.
i\sin\frac{\phi_{\mathbf{q}}-\phi_{\mathbf{q}\prime}}{2}\left(
\delta_{\mathbf{p,q}}\xi_{\mathbf{q}s}^{\dag}\xi_{\mathbf{q}^{\prime}
,-s}-\delta_{\mathbf{p,q}^{\prime}}\xi_{\mathbf{q},-s}^{\dag}\xi
_{\mathbf{q}^{\prime}\mathbf{,}s}\right)  \right]. \label{norcom}
\end{align}
Averaging this expression over QSO gives
$\mathcal{J}^{(1)}_{\mathbf{p}}\left[f_s \right ] = 0$. Then,
performing the second commutation using Eqs. (\ref{GotL}) and
(\ref{norcom}), after straightforward calculations we obtain the
second-order collision integral for $f_{\mathbf{p}s}$
\begin{align}
\mathcal{J}_{\mathbf{p}}^{(2)}\left[ f_{s}\right]   &  = S^{-2}
\sum_{\mathbf{q} }\left\vert V\left(  \mathbf{p-q}\right)
\right\vert ^{2}\left\{ s\sin\left(
\phi_{\mathbf{p}}-\phi_{\mathbf{q}}\right) \operatorname{Re}\left(
g_{\mathbf{q}}\right)  \left( \frac{1}{\epsilon
_{p}+\epsilon_{q}}+\frac{1}{\epsilon_{q}-\epsilon_{p}}\right) \right. \nonumber \\
&  - \left. \pi\left[
2\cos^{2}\frac{\phi_{\mathbf{p}}-\phi_{\mathbf{q}}}{2}\left(
f_{\mathbf{p} s}-f_{\mathbf{q}s}\right)  +s\sin\left(
\phi_{\mathbf{p}}-\phi_{\mathbf{q} }\right)
\operatorname{Im}\left(  g_{\mathbf{q}}\right)  \right]
\delta\left(  \epsilon_{p}-\epsilon_{q}\right) \right\}
\label{normal} .
\end{align}
Further we have
\begin{align}
\left[
\xi_{\mathbf{p}s}^{\dag}\xi_{\mathbf{p},-s},\mathcal{H}_{\text{int}
}\right]   &  = S^{-1}\sum_{\mathbf{qq}^{\prime}}V\left(
\mathbf{q-q}^{\prime }\right)  \left[
\cos\frac{\phi_{\mathbf{q}}-\phi_{\mathbf{q}^{\prime}}} {2}\left(
\delta_{\mathbf{pq}}-\delta_{\mathbf{pq}^{\prime}}\right)
\xi_{\mathbf{q}s}^{\dag}\xi_{\mathbf{q}^{\prime},-s}\right. \nonumber \\
&  +i\left.
\sin\frac{\phi_{\mathbf{q}}-\phi_{\mathbf{q}\prime}}{2}\left(
\delta_{\mathbf{pq}}\xi_{\mathbf{q}s}^{\dag}\xi_{\mathbf{q}^{\prime}s}
-\delta_{\mathbf{pq}^{\prime}}\xi_{\mathbf{q,-}s}^{\dag}\xi_{\mathbf{q}
^{\prime},-s}\right)  \right]. \label{ancom}
\end{align}
Making the second commutation with the use of Eqs. (\ref{GotL})
and (\ref{ancom}), after straightforward calculations we obtain
the second-order collision integral for $g_{\mathbf{p}s}$
\begin{align}
\mathcal{J}_{\mathbf{p}}^{(2)}\left[  g_{s}\right]   & =
-S^{-2}\sum_{\mathbf{qq}^{\prime }}\left\vert V\left(
\mathbf{p-q}\right) \right\vert ^{2}\left\{  2\cos
^{2}\frac{\phi_{\mathbf{p}}-\phi_{\mathbf{q}}}{2}\right.  \times \nonumber \\
&  \left[  \left(  g_{\mathbf{p}s}-g_{\mathbf{q}s}\right)
\pi\delta\left( \epsilon_{p}-\epsilon_{q}\right)
+is\frac{g_{\mathbf{p}s}+g_{\mathbf{q}s}
}{\epsilon_{q}-\epsilon_{p}}\right]  -\frac{i}{2}\sin\left(
\phi_{\mathbf{p}
}-\phi_{\mathbf{q}}\right)  \times \nonumber \\
&  \left(  f_{\mathbf{q}s}+f_{\mathbf{q,-}s}\right)  \left[
\pi\delta\left( \epsilon_{p}-\epsilon_{q}\right)
-is\frac{1}{\epsilon_{q}-\epsilon_{p}
}\right]  + \nonumber \\
&  \left.  \frac{s}{2}\sin\left(
\phi_{\mathbf{p}}-\phi_{\mathbf{q}}\right)
\frac{f_{\mathbf{q}s}-f_{\mathbf{q},-s}}{\epsilon_{p}+\epsilon_{q}}
-2is\sin^{2}\frac{\phi_{\mathbf{p}}-\phi_{\mathbf{q}}}{2}\frac{g_{\mathbf{p}
s}+g_{\mathbf{q}s}^{\ast}}{\epsilon_{p}+\epsilon_{q}}\right\}.
\label{abnormal}
\end{align}
This equation describes the Zitterbewegung effects, that is,
creation of electron-hole pairs during the charge carrier
propagation. Note that in both Eq. (\ref{normal}) and
(\ref{abnormal}) the configurational average of the potential
Fourier transforms squared is implied.

Putting together Eqs.(\ref{field}),(\ref{normal}) and
({\ref{abnormal}) we can write the final set of GKE for the
``normal'' and ``anomalous'' distribution functions. It is more
convenient, however, to transform these GKE to a system of
equations for the functions
\begin{equation}
D_{\mathbf{p}}=\sum_{s = \pm1}f_{\mathbf{p}s}-1=n_{\mathbf{p}
}-p_{\mathbf{p}},\;N_{\mathbf{p}}=\sum_{s=\pm1}sf_{\mathbf{p}s}
+1=n_{\mathbf{p}}+p_{\mathbf{p}}
\end{equation}
and for $g_{\mathbf{p}}$. Let us remind that the latter and
$N_{\mathbf{p}}$ define the average current, see Eq.
(\ref{avcurrent}). We have for $D_{\mathbf{p}}$
\begin{equation}
\frac{\partial D_{\mathbf{p}}}{\partial t}+eE\frac{\partial D
_{\mathbf{p}}}{\partial
p_{x}}=-\frac{2\pi}{S^2}\sum_{\mathbf{q}}\left\vert V\left(
\mathbf{p-q}\right)  \right\vert
^{2}\cos^{2}\frac{\phi_{\mathbf{p}} -\phi_{\mathbf{q}}}{2}\left(
D_{\mathbf{p}}- D_{\mathbf{q}}\right) \delta\left(
\epsilon_{p}-\epsilon_{q}\right),  \label{b1}
\end{equation}
which does not involve $g_{\mathbf{p}}$ at all, while the equation
for $N_{\mathbf{p}}$ reads
\begin{align}
& \frac{\partial N_{\mathbf{p}}}{\partial t}+ eE\frac{\partial
N_{\mathbf{p}} }{\partial
p_{x}}-\frac{2eE\sin\phi_{\mathbf{p}}}{p}\operatorname{Im}
g_{\mathbf{p}} \nonumber &  \\
& = \frac{2\pi}{S^2}\sum_{\mathbf{q}}\left\vert V\left(
\mathbf{p-q}\right) \right\vert ^{2}\left\{ \pi^{-1}\sin\left(
\phi_{\mathbf{p}}-\phi_{\mathbf{q}}\right) \operatorname{Re}\left(
g_{\mathbf{q}}\right)  \left( \frac{1}{\epsilon
_{q}+\epsilon_{p}}+\frac{1}{\epsilon_{q}-\epsilon_{p}}\right)
  \right. \nonumber \\
&  -\left. \left[
\cos^{2}\frac{\phi_{\mathbf{p}}-\phi_{\mathbf{q}}} {2}\left(
N_{\mathbf{p}}-N_{\mathbf{q}}\right)  +\sin\left(  \phi
_{\mathbf{p}}-\phi_{\mathbf{q}}\right)  \operatorname{Im}\left(
g_{\mathbf{q}}\right)  \right]  \delta\left( \epsilon_{p}-\epsilon
_{q}\right)\right\} .  \label{b2}
\end{align}
Finally, the equation for the complex function $g_{\mathbf{p}}$
proves the following
\begin{align}
& \frac{\partial g_{\mathbf{p}}}{\partial t}-2ivp\,g_{\mathbf{p}}
+ eE\frac{\partial\,g_{\mathbf{p}}}{\partial
p_{x}}+i\frac{eE}{2p}\left( N_{\mathbf{p}}-1\right)
\sin\phi_{\mathbf{p}} = \nonumber \\
&  -\frac{\pi}{S^2}\sum_{\mathbf{q}}\left\vert V\left(
\mathbf{p-q}\right) \right\vert ^{2}\left\{ -\frac{i}{2}\sin\left(
\phi_{\mathbf{p}}-\phi_{\mathbf{q} }\right) D_{\mathbf{q}}\left[
\delta\left(  \epsilon_{p} -\epsilon_{q}\right)
+\frac{i}{\pi}\frac{1}{\epsilon_{p}-\epsilon_{q}}\right] \right.
\nonumber \\ &  +
2\cos^{2}\frac{\phi_{\mathbf{p}}-\phi_{\mathbf{q}}}{2}\left[
\left( g_{\mathbf{p}}-g_{\mathbf{q}}\right)  \delta\left(
\epsilon_{p} -\epsilon_{q}\right) +
\frac{i}{\pi}\frac{g_{\mathbf{p}}+g_{\mathbf{q}}}{\epsilon
_{q}-\epsilon_{p}}\right]  \nonumber \\
&  +\left.
\frac{1}{2\pi}\frac{N_{\mathbf{q}}}{\epsilon_{p}+\epsilon_{q}}\sin\left(
\phi_{\mathbf{p}}-\phi_{\mathbf{q}}\right)
-\frac{2i}{\pi}\frac{g_{\mathbf{p}}+g_{\mathbf{q}
}^{\ast}}{\epsilon_{p}+\epsilon_{q}}\sin^{2}\frac
{\phi_{\mathbf{p}}-\phi_{\mathbf{q}}}{2}\right\} . \label{b3}
\end{align}

\section{THE LINEAR RESPONSE REGIME IN ELECTRIC FIELD}

In general, Eqs. (\ref{b1}) - (\ref{b3}) are quite complicated.
Further we will consider only the regime of linear response, that
is, the case of weak electric field. We will restrict ourselves also
by the case of stationary field and neglect its effect on the
collision integral. It can be shown that linear in $E$ corrections
to $\mathcal{J}^{(2)}_{\mathbf{p}}$ restore some second order terms
of the perturbation expansion of the exact field term considered in
the Luttinger-Kohn formalism \cite{kohn}, which lies out of our
scope.

At $E = 0$, like the classical kinetic equation, Eqs. (\ref{b1}) -
(\ref{b3}) have equilibrium solution. In our case it is three
arbitrary function of the energy $\epsilon_p = vp$. To develop
meaningful linearization of GKE and the current for small $E$, we
assume, following Ref. \onlinecite{kohn}, presence of a thermostat
which role is only to establish true equilibrium with a temperature
$T$. Under this condition the equilibrium distribution functions
$n_{0\mathbf{p}}$ and $p_{0\mathbf{p}}$ become, of course, the
electrons and holes Fermi functions, respectively, with unique
chemical potential $\mu$. Now let us consider the linearization of
GKE derived above in detail.

\subsection{Linearized equation for $D_{\mathbf{p}}$}

Replacing in the field term of Eq.(\ref{b1}) the distribution
functions by their equilibrium value and linearizing the
corresponding collision integral, we obtain the following equation
\begin{equation}
eE\frac{\partial D_{0}\left(  \epsilon_{p}\right)  }{\partial
p_{x}} =-\frac{2\pi}{S^2}\sum_{\mathbf{q}}\left\vert V\left(
\mathbf{p-q}\right)  \right\vert
^{2}\cos^{2}\frac{\phi_{\mathbf{p}}-\phi_{\mathbf{q}}}{2}\left(
\delta D_{\mathbf{p}}- \delta D_{\mathbf{q}}\right) \delta\left(
\epsilon _{p}-\epsilon_{q}\right), \label{lineq-Delta}
\end{equation}
where
\begin{equation}
 D_{0}\left(  \epsilon\right)
=\frac{1}{e^{\frac{\epsilon-\mu}{T}}
+1}-\frac{1}{e^{\frac{\epsilon+\mu}{T}}+1}, \label{Delta_0}
\end{equation}
Eq. (\ref{lineq-Delta}) is quite similar to the classical kinetic
equation and so is routinely solved exactly. The solution reads
\begin{equation}
\delta D_{\mathbf{p}}= -eEv\tau\left( \epsilon_{p}\right)
\frac{\partial D_{0}\left( \epsilon_{p}\right)
}{\partial\epsilon_{p}}\cos\phi_{\mathbf{p}}, \label{Delta_p}
\end{equation}
$\tau\left(  \epsilon_{p}\right)$ being standard elastic transport
relaxation time given by
\begin{align}
\frac{1}{\tau\left(  \epsilon_{p}\right)  }  & =
\frac{\pi}{S^2}\sum_{\mathbf{q} }\left\vert V\left(
\mathbf{p-q}\right) \right\vert ^{2} \sin^{2}\left(
\phi_{\mathbf{p}}-\phi_{\mathbf{q}}\right) \delta\left(
\epsilon_{p}-\epsilon_{q}\right)  =\nonumber\\
&  =\frac{\pi \epsilon_p}{\left( 2\pi v\right)
^{2}}N_{\text{imp}}\int_{0}^{2\pi}\left\vert U\left(
2p\sin\frac{\phi}{2}\right) \right\vert ^{2}\sin^{2}\phi d\phi ,
\label{invtau}
\end{align}
where $N_{\text{imp}} = c/\Omega$ is the impurity concentration
per the area unit, $c$ and $\Omega$ being atomic fraction of
impurities and the graphene crystal cell area, respectively, and
$U\left (\vert \mathbf{q} \vert \right )$ is the Fourier transform
of one-impurity potential. In deriving Eq. (\ref{invtau}) and what
follows we adopted that $c \ll 1$. The factor $\sin^2 {\phi}$
instead of standard \cite{ziman} one $1-\cos{\phi}$ is a
consequence of the chiral character of charge carriers which leads
to the suppressions of back scattering \cite{r4,shon}.

\subsection{Linearized equations for $N_{\mathbf{p}}$ and $g_{\mathbf{p}}$}

Replacing in the field terms of Eqs. (\ref{b2}) and (\ref{b3}) the
distribution functions by their equilibrium value and linearizing
the corresponding collision integrals, we arrive at the following
system of coupled equations
\begin{align}
&  eE\frac{\partial N_{0}\left(  \epsilon_{p}\right)  }{\partial
p_{x}}
-\frac{2eE\sin\phi_{\mathbf{q}}}{q}\operatorname{Im}\,g_{0}\left(
\epsilon_{p}\right)  = \nonumber\\
&   \frac{2\pi}{S^2} \sum_{\mathbf{q}}\left\vert V\left(
\mathbf{p-q}\right) \right\vert ^{2}\left\{ \sin\left(
\phi_{\mathbf{p}}-\phi_{\mathbf{q}}\right) \left(
\frac{1}{\epsilon_{q} +\epsilon_{p}}+
\frac{1}{\pi}\frac{1}{\epsilon_{q}-\epsilon_{p}}\right)
\operatorname{Re}\delta g_{\mathbf{q}}
 \right. \nonumber\\
&  \left. - \left[
\cos^{2}\frac{\phi_{\mathbf{p}}-\phi_{\mathbf{q}}} {2}\left(
\delta N_{\mathbf{p}}- \delta N_{\mathbf{q}}\right) +\sin\left(
\phi _{\mathbf{p}}-\phi_{\mathbf{q}}\right)
\operatorname{Im}\delta g_{\mathbf{q} }\right]  \delta\left(
\epsilon_{p}-\epsilon_{q}\right) \right\} \label{eq-N-g}
\end{align}
and
\begin{align}
&  -2i\epsilon_{p} \delta g_{\mathbf{p}}+eE\frac{\partial
g_{0}\left( \epsilon _{p}\right) }{\partial
p_{x}}+i\frac{eE\sin\phi_{\mathbf{p}}}{2p}\left[ N_{0}\left
(\epsilon_p \right ) -1\right]
=\nonumber\\
&  - \frac{\pi}{S^2}\sum_{\mathbf{q}}\left\vert V\left(
\mathbf{p-q}\right) \right\vert ^{2}\left\{ -\frac{i}{2}\delta
D_{\mathbf{q}}\left[  \delta\left(  \epsilon_{p}
-\epsilon_{q}\right) +\frac{i}{\pi}\frac{1}{
\epsilon_{p}-\epsilon_{q} }\right]\sin\left(
\phi_{\mathbf{p}}-\phi_{\mathbf{q} }\right)  \right.
\nonumber\\
&  +2 \left[ \left( \delta g_{\mathbf{p}}-\delta
g_{\mathbf{q}}\right) \delta\left(
\epsilon_{p}-\epsilon_{q}\right)  + \frac{i}{\pi}\frac{\delta
g_{\mathbf{p}}+\delta
g_{\mathbf{q}}}{\epsilon_{q}-\epsilon_{p}}\right]\cos^{2}
\frac{\phi_{\mathbf{p}}-\phi_{\mathbf{q}}}{2}
\nonumber\\
&  +\left.  \frac{1}{2\pi} \frac{\delta
N_{\mathbf{q}}}{\epsilon_{p}+\epsilon_{q}} \sin\left(
\phi_{\mathbf{p}}-\phi_{\mathbf{q}}\right)-\frac{2i}{\pi}\frac{\delta
g_{\mathbf{p}}+\delta
g_{\mathbf{q}}^{\ast}}{\epsilon_{p}+\epsilon_{q}}\sin^{2}\frac
{\phi_{\mathbf{p}}-\phi_{\mathbf{q}}}{2}\right\}  , \label{eq-g-N}
\end{align}
where
\begin{equation}
N_{0}\left(  \epsilon\right)
=\frac{1}{e^{\frac{\epsilon-\mu}{T}}+1}+\frac
{1}{e^{\frac{\epsilon+\mu}{T}}+1}, \label{N_0}
\end{equation}
and $g_{0}\left(  \epsilon_{p}\right) $ is an equilibrium
``anomalous'' distribution function satisfying the equation
\begin{align*}
&  2i\epsilon_p g_{0}\left(  \epsilon_{p}\right) =
\frac{\pi}{S^2}\sum_{\mathbf{q}}\left\vert V\left(
\mathbf{p-q}\right) \right\vert ^{2}\left\{ -\frac{i}{2}
 D_{0}\left( \epsilon _{q}\right)  \left[  \delta\left(
\epsilon_{p}-\epsilon_{q}\right)
+\frac{i}{\pi}\frac{1}{\epsilon_{p}-\epsilon_{q}}\right]
\sin\left(
\phi_{\mathbf{p}}-\phi_{\mathbf{q}}\right) \right.  +\nonumber\\
&  \left. \frac{2i}{\pi}\left [\frac {g_{0}\left(
\epsilon_{p}\right) +g_{0}\left( \epsilon_{q}\right)
}{\epsilon_{q}-\epsilon_{p}}\cos^{2}\frac{\phi_{\mathbf{p}}-\phi_{\mathbf{q}}}{2}
- \frac{g_{0}\left( \epsilon_{p}\right) +g_{0}^{\ast}\left(
\epsilon_{q}\right)
}{\epsilon_{p}+\epsilon_{q}}\sin^{2}\frac{\phi_{\mathbf{p}}-\phi
_{\mathbf{q}}}{2}\right ] +\frac{1}{2\pi}\frac{N_{0}\left(
\epsilon_{q}\right)  } {\epsilon_{p}+\epsilon_{q}} \sin\left(
\phi_{\mathbf{p}} -\phi_{\mathbf{q}}\right) \right\}  .
\end{align*}
Because
\begin{align*}
\int_{0}^{2\pi}\left\vert V\left(  \mathbf{p-q}\right) \right\vert
^{2} \sin\left( \phi_{\mathbf{p}}-\phi_{\mathbf{q}}\right)
d\phi_{\mathbf{q}}  \propto  \int_{0}^{2\pi}\left\vert U\left(
\sqrt{p^{2}-2pq\cos\chi+q^{2}}\right) \right\vert ^{2}\sin\chi
d\chi\equiv 0,
\end{align*}
the terms containing $ D_{0}\left( \epsilon _{q}\right)  $ and
$N_{0}\left(  \epsilon_{q}\right) $ in the above equation for
$g_{0}\left( \epsilon_{p}\right) $ give zero contributions. As a
result, if even non-zero $g_{0}\left( \epsilon\right)$ is purely
real. In contrast with $N_{0}\left (\epsilon \right )$ and
$D_{0}\left (\epsilon \right )$, this real function has no influence
on the non-equilibrium $\delta N_{\mathbf{p}}$ and $\delta
g_{\mathbf{p}}$ and so drops out of our treatment.

Let us search the functions $\delta N_{\mathbf{p}}$ and $\delta
g_{\mathbf{p}}$ in the form
\begin{equation}
\delta N_{\mathbf{p}}= -eE\nu\left( \epsilon _{p}\right)
\cos\phi_{\mathbf{p}} \label{deltaN_p}
\end{equation}
and
\begin{equation}
\delta\,g_{\mathbf{p}}=eE\gamma\left(  \epsilon_{p}\right)  \sin
\phi_{\mathbf{p}}, \label{deltag}
\end{equation}
respectively. Substituting these forms into expression for the
average current (\ref{jxxx}) yields the following expression for
the conductivity
\begin{equation}
\sigma = \frac{e^{2}}{2\pi v}\int_{0}^{\infty}\left[
\operatorname{Im} \gamma\left(  \epsilon\right) -\frac{\nu\left(
\epsilon \right)}{2} \right]  \epsilon d\epsilon. \label{sigmaxx}
\end{equation}
Substitute further Eqs.(\ref{deltaN_p}) and (\ref{deltag}) into
Eq.(\ref{eq-N-g}). Taking into account that
$\operatorname{Im}g_{0}\left( \epsilon_{p}\right)  = 0$, we manage
to show that all non-zeroing terms of the resulting equation
contain common factor $\cos\phi_{\mathbf{p}}$. Dividing by this
factor we get the first linear-response GKE
\begin{equation}
v\frac{\partial N_{0}\left(  \epsilon_{p}\right)
}{\partial\epsilon_{p} }=\frac{\nu\left(  \epsilon_{p}\right)
+2\operatorname{Im}\gamma\left( \epsilon_{p}\right)  }{\tau\left(
\epsilon_{p}\right)  }-2\int_{0}^{\infty }\Gamma_{0}\left(
p,q\right)  \left(  \frac{1}{q+p}+\frac{1}{q-p}\right)
\operatorname{Re}\gamma\left(  \epsilon_{q}\right)  dq,
\label{eq-nu}
\end{equation}
where the kernel in the integral term is given by
\begin{equation}
\Gamma_{0}\left(  p,q\right)  =\frac{\epsilon_q}{\left(  2\pi
v\right) ^{2}} N_{\text{imp}} \int_{0}^{2\pi}\left\vert U\left(
\sqrt{p^{2}-2pq\cos\phi+q^{2}}\right) \right\vert ^{2}\sin^{2}\phi
d\phi\label{Gamma_0}
\end{equation}
It is expedient to note that
\begin{equation}
 \Gamma_{0}\left(  p,p\right)
=\frac{1}{\pi \tau \left( \epsilon_{p} \right) } .
\label{Gamma0-tau}
\end{equation}
Performing quite similar transformations in the ``anomalous''
kinetic equation, Eq.(\ref{eq-g-N}), we find that all non-zeroing
terms of the resulting equation are  proportional to
$\sin\phi_{\mathbf{p}}$. Cancelling this common factor, we obtain
the second linear-response GKE
\begin{align}
&  -2i\epsilon_{p}\gamma\left(  \epsilon_{p}\right)  =\frac{iv}{2}\left[
\frac{1-N_{0}\left(  \epsilon_{p}\right)  }{\epsilon_{p}}-\frac{\partial
 D_{0}\left(  \epsilon_{p}\right)  }{\partial\epsilon_{p}}\right]
\nonumber\\
&  -\frac{ N_{\text{imp}}}{\left(  2\pi\right) ^{2}}\int\left\vert
U\left( \mathbf{p-q}\right)  \right\vert ^{2}\left\{
-\frac{1}{2}\sin^{2}\left(
\phi_{\mathbf{p}}-\phi_{\mathbf{q}}\right)  \left[ \frac{\partial
D_{0}\left(  \epsilon_{q}\right)
}{\partial\epsilon_{q}}\frac{v\tau\left( \epsilon_{q}\right)
}{\epsilon_{p}-\epsilon_{q}}+\frac{\nu\left(
\epsilon_{p}\right)  }{\epsilon_{p}+\epsilon_{q}}\right]  \right. \nonumber\\
&  +\left[  1+\cos\left(  \phi_{\mathbf{q}}-\phi_{\mathbf{p}}\right)  \right]
\left[  \gamma\left(  \epsilon_{p}\right)  \left(  1-\cos\left(
\phi_{\mathbf{q}}-\phi_{\mathbf{p}}\right)  \right)  \pi\delta\left(
\epsilon_{p}-\epsilon_{q}\right)  \right. \nonumber\\
&  +\left.  i\frac{\gamma\left(  \epsilon_{p}\right)  +\gamma\left(
\epsilon_{q}\right)  \cos\left(  \phi_{\mathbf{q}}-\phi_{\mathbf{p}}\right)
}{\epsilon_{q}-\epsilon_{p}}\right] \nonumber\\
-  &  i\left.  \left[  1-\cos\left(
\phi_{\mathbf{q}}-\phi_{\mathbf{p} }\right)  \right]
\frac{\gamma\left(  \epsilon_{p}\right)  +\gamma^{\ast }\left(
\epsilon_{q}\right)  \cos\left( \phi_{\mathbf{q}}-\phi_{\mathbf{p}
}\right)  }{\epsilon_{p}+\epsilon_{q}}\right\}  d^2 q.
\label{eq-gam}
\end{align}
\subsection{Linear-response GKE in the energy variable}
In what follows we will make overall use of the energy variable
$\epsilon = vp$. Let us summarize linear-response GK, i.e Eqs.
(\ref{eq-nu}) and (\ref{eq-gam}), in terms of $\epsilon$. We have
\begin{equation}
v\frac{\partial N_{0}\left(  \epsilon\right)
}{\partial\epsilon}=\frac {\nu\left(  \epsilon\right)
}{\tau\left(  \epsilon\right)  }+\frac
{2\operatorname{Im}\gamma\left(  \epsilon\right)  }{\tau\left(
\epsilon \right)  }-2\int_{0}^{\infty}\widetilde{\Gamma}_{0}\left(
\epsilon ,\omega\right)  \left(
\frac{1}{\omega+\epsilon}+\frac{1}{\omega-\epsilon }\right)
\operatorname{Re}\gamma\left(  \omega\right)
d\omega\label{eq-nu-e}
\end{equation}
and
\begin{align}
&  -2i\epsilon\gamma\left(  \epsilon\right)  =\frac{iv}{2}\left[
\frac{1-N_{0}\left(  \epsilon\right)  }{\epsilon}-\frac{\partial
D_{0}\left(  \epsilon\right)  }{\partial\epsilon}\right]
+\frac{1}{2}\int _{0}^{\infty}\widetilde{\Gamma}_{0}\left(
\epsilon,\omega\right)  \left[ \frac{\nu\left(  \omega\right)
}{\epsilon+\omega}+\frac{\partial D_{0}\left(  \omega\right)
}{\partial\omega}\frac{v\tau\left(  \omega\right)
}{\epsilon-\omega}\right]  d\omega  \nonumber \\
&  -\frac{\gamma\left(  \epsilon\right)  }{\tau\left(
\epsilon\right) }+i\Delta\left(  \epsilon\right)  \gamma\left(
\epsilon\right)  +i\int _{0}^{\infty}\left[
\frac{\widetilde{\Gamma}_{+}\left(  \epsilon ,\omega\right)
}{\epsilon-\omega}\gamma\left(  \omega\right)  +\frac
{\widetilde{\Gamma}_{-}\left(  \epsilon,\omega\right)
}{\epsilon+\omega }\gamma^{\ast}\left(  \omega\right)  \right]
d\omega,
\end{align}
where the kernels are given by
\begin{align}
& \widetilde{\Gamma}_{0}\left( \epsilon,\omega\right)
=\frac{\omega}{\left(  2\pi v\right)
^{2}}N_{\text{imp}}\int_{0}^{2\pi}\left\vert U\left(
\frac{1}{v}\sqrt
{\epsilon^{2}-2\epsilon\omega\cos\phi+\omega^{2}}\right)
\right\vert ^{2} \sin^{2}\phi d\phi\equiv \Gamma_{0}\left(
\frac{\epsilon}{v},\frac{\omega}{v}\right), \nonumber \\
& \widetilde{\Gamma}_{\pm}\left(  \epsilon,\omega\right) =
\frac{\omega }{\left(  2\pi v\right)
^{2}}N_{\text{imp}}\int_{0}^{2\pi}\left\vert U\left(  \frac{1}
{v}\sqrt{\epsilon^{2}-2\epsilon\omega\cos\phi+\omega^{2}}\right)
\right\vert ^{2}\left( 1 \pm\cos\phi\right)\cos\phi d\phi
\label{Gamma}
\end{align}
and, using these kernels, two-particle energy shift by
\begin{equation}
\Delta\left(  \epsilon\right)  =\int_{0}^{\infty}\left[ \frac{
\widetilde{\Gamma}_{0}\left(
\epsilon,\omega\right)+\widetilde{\Gamma}_{+}\left(
\epsilon,\omega\right) }{\epsilon-\omega}+\frac{\widetilde{\Gamma}
_{0}\left( \epsilon,\omega\right)-\widetilde{\Gamma}_{-}\left(
\epsilon,\omega\right) }{\omega+\epsilon}\right]  d\omega.
\label{del}
\end{equation}

Let us introduce dimensionless kernels $\Phi_{\alpha}\left(
\epsilon,\omega\right)$, $\alpha = 0,\pm$ by the following
identity relations
\begin{equation}
\widetilde{\Gamma}_{0}\left( \epsilon,\omega\right) \equiv
\omega\Phi_{0}\left( \epsilon,\omega\right), \hspace{0.3 cm}
\widetilde{\Gamma}_{\pm}\left(  \epsilon,\omega\right) \equiv
\pm\omega\Phi_{\pm }\left( \epsilon,\omega\right),
\end{equation}
with which and Eq. (\ref{Gamma}) the formulas of these new kernels
being straightforward. Note that, by construction of the kinetic
equations, $ \left\vert \Phi_{\alpha }\left(
\epsilon,\omega\right) \right\vert \ll1$. Making use of the
kernels $\Phi_{\alpha}\left(\epsilon,\omega\right) $ and of new
functions of the energy defined by
\begin{equation}
\epsilon\nu\left(  \epsilon\right)  =  v \left [2f_{0}\left(
\epsilon\right) + \frac{1}{\pi \Phi_{0}\left (\epsilon,\epsilon
\right )}\frac{\partial N_{0}\left (\epsilon \right )}{\partial
\epsilon} \right ] ,\;\epsilon\gamma\left( \epsilon\right) =
vf\left( \epsilon\right) ,\;f=f_{1}+if_{2}\label{newf}
\end{equation}
we arrive at a modified conductivity expression and two coupled
singular integral equations for $f_{0}\left( \epsilon\right)$ and
$f\left( \epsilon\right)$. We have
\begin{equation}
\sigma =
\sigma_{\text{B}}+\frac{e^{2}}{2\pi}\int_{0}^{\infty}\left[
f_{2}\left( \epsilon\right)  -f_{0}\left(  \epsilon\right) \right]
d\epsilon, \label{sigmaxx1}
\end{equation}
where
\begin{equation}
\sigma_{\text{B}} = - \frac{e^2}{\left( 2\pi \right
)^2}\int_{0}^{\infty}\frac{1}{\Phi_{0}\left (\epsilon,\epsilon
\right )} \frac{\partial N_{0} \left (\epsilon \right)}{\partial
\epsilon} d\epsilon
\end{equation}
is the classical Boltzmann conductivity \cite{ziman}, with the Born
impurity scattering cross-section modified due to chirality as noted
above, in which of electrons and holes contribute additively, while
the additional non-classical term is due to the Zitterbewegung. The
integral equations are as follows. The first one is real and
homogenous
\begin{equation}
\pi\Phi_{0}\left(  \epsilon,\epsilon\right)  \left[  f_{2}\left(
\epsilon\right)  +f_{0}\left(  \epsilon\right)  \right]
-\int_{0}^{\infty }\Phi_{0}\left(  \epsilon,\omega\right)  \left(
\frac{1}{\omega+\epsilon }+\frac{1}{\omega-\epsilon}\right)
f_{1}\left(  \omega\right)  d\omega = 0 \label{eq1-gen}
\end{equation}
and the second equation is complex and inhomogeneous
\begin{align}
&  \left[  \Lambda\left(  \epsilon\right)  +i\pi\Phi_{0}\left(  \epsilon
,\epsilon\right)  \right]  f\left(  \epsilon\right)  -\int_{0}^{\infty}\left[
\frac{\Phi_{+}\left(  \epsilon,\omega\right)  }{\omega-\epsilon}f\left(
\omega\right)  +\frac{\Phi_{-}\left(  \epsilon,\omega\right)  }{\omega
+\epsilon}f^{\ast}\left(  \omega\right)  \right]  d\omega\nonumber\\
&  -i\int_{0}^{\infty}\frac{\Phi_{0}\left(  \epsilon,\omega\right)
} {\epsilon+\omega}f_{0}\left(  \omega\right)  d\omega = F\left(
\epsilon\right) ,\label{eq2-gen}
\end{align}
in which
\begin{equation}
\Lambda\left( \epsilon\right)  =2 + \epsilon^{-1} \Delta\left(
\epsilon \right) \label{Lambda}
\end{equation}
and $F\left( \epsilon\right) = F_{1}\left( \epsilon\right)
+iF_{2}\left( \epsilon\right)$, where
\begin{align}
& F_{1}\left(  \epsilon\right) =-\frac{1-N_{0}\left(
\epsilon\right)
}{2\epsilon}+\frac{1}{2}\frac{\partial D_{0}\left( \epsilon\right)}{\partial\epsilon}, \nonumber\\
F_{2}\left(  \epsilon\right)   &
=\frac{1}{2\pi}\int_{0}^{\infty}\frac {\Phi_{0}\left(
\epsilon,\omega\right)  }{\Phi_{0}\left(  \omega ,\omega\right)
}\left [\frac{1}{\omega+\epsilon}\frac{\partial N_{0}\left(
\omega\right)
}{\partial\omega}-\frac{1}{\omega-\epsilon}\frac{\partial
D_{0}\left( \omega\right) }{\partial\omega} \right ]d\omega.
\label{F}
\end{align}
Eq.(\ref{eq2-gen}) is, in turn, equivalent to two real equations
for $f_{1}$ and $f_{2}$
\begin{equation}
\Lambda\left(  \epsilon\right)  f_{1}\left(  \epsilon\right)
-\pi\Phi _{0}\left(  \epsilon,\epsilon\right)  f_{2}\left(
\epsilon\right)  -\int _{0}^{\infty}\left[  \frac{\Phi_{+}\left(
\epsilon,\omega\right)  } {\omega-\epsilon}+\frac{\Phi_{-}\left(
\epsilon,\omega\right)  } {\omega+\epsilon}\right]  f_{1}\left(
\omega\right)  d\omega=F_{1}\left( \epsilon\right)  \label{eq2-1}
\end{equation}
and
\begin{align}
\pi\Phi_{0}\left(  \epsilon,\epsilon\right)  f_{1}\left(
\epsilon\right)  + &  \Lambda\left(  \epsilon\right)  f_{2}\left(
\epsilon\right)  -\int _{0}^{\infty}\left[  \frac{\Phi_{+}\left(
\epsilon,\omega\right)  } {\omega-\epsilon}-\frac{\Phi_{-}\left(
\epsilon,\omega\right)  }
{\omega+\epsilon}\right]  f_{2}\left(  \omega\right)  d\omega\nonumber\\
&  -\int_{0}^{\infty}\frac{\Phi_{0}\left(  \epsilon,\omega\right)
} {\omega+\epsilon}f_{0}\left(  \omega\right)  d\omega=F_{2}\left(
\epsilon\right),  \label{eq2-2}
\end{align}
respectively.

\section{A short-range impurity
potential model}

In this section we solve the system of the integral equations
derived above, see Eqs. (\ref{eq1-gen}) and (\ref{eq2-gen}), or
the real equivalents of the latter - Eqs. (\ref{eq2-1}) and
(\ref{eq2-1}), and then calculate the conductivity using Eq.
(\ref{sigmaxx1}) for a weak extremely short-range impurity
potential.

\subsection{Formulation of the model}

Let us consider a zero-range impurity potential, which we define
by $U\left( \mathbf{r} \right) = U_{0}\Omega \delta \left(
\mathbf{r} \right) $, where the parameter $U_0$ has dimension of
energy. For this potential, $U\left( \mathbf{q} \right) = \Omega
U_{0}$ so the kernels $\Phi_{\alpha}$ become independent of their
energy arguments and all equal
\begin{equation}
\Phi_{\alpha}\left (\epsilon, \omega \right ) = \Phi =
\frac{cU_{0}^{2}\Omega }{4 \pi v^2 \hbar^2}
\end{equation}
Here where we have restored the Planck constant $\hbar$ to stress
that $\Phi$ is dimensionless. This approximation poses no formal
problem as regards the integral terms in Eqs. (\ref{eq1-gen}) and
(\ref{eq2-gen}) but $\Lambda\left (\epsilon \right )$ acquires an
ultraviolet logarithmic divergence. Thus some ultraviolet cut off
procedure should be introduced. We define $\Phi_{\alpha}\left(
\epsilon,\omega\right) = \Phi$ at
$0\leq\epsilon\leq\epsilon_{c},\,0\leq\omega\leq\epsilon_{c}$ and
zero otherwise. Then the simple calculation using Eqs.
(\ref{Lambda}) and (\ref{del}) yields
\begin{equation}
\Lambda\left(  \epsilon\right) = 2 - 4\Phi\int_{0}^{\epsilon_{c}
}\frac{\omega}{\omega^{2}-\epsilon^{2}}d\omega = 2 \left [1 -
\Phi\ln\left (\frac{\epsilon^{2}}
{\epsilon_{c}^{2}-\epsilon^{2}}\right )\right ] \label{DDelta}
\end{equation}
at $\epsilon < \epsilon_{c}$ and $\Lambda\left(  \epsilon\right) = 2
$ otherwise. This function has infinity breakpoint at $\epsilon =
\epsilon_c$, which has no physical meaning. Analysis of a general
case with a finite-range potential shows that the model is
reasonable assuming that we are interested only in small enough
energies in comparison with $\epsilon_c = \hbar v/r_0$ where $r_0$
is a characteristic radius of the potential. On the other hand, the
noted mathematical property of the model $\Lambda\left(
\epsilon\right) $ allows us to solve a part of the obtained singular
integral equation exactly, see Appendix.

\subsection{The model integral equations and their solution}

Using the introduced model we obtain from the following system of
singular integral equations for $\epsilon\leq\epsilon_{c}$
\begin{equation}
  f_{2}\left(  \epsilon\right)  +f_{0}\left( \epsilon\right)
 -\frac{2}{\pi}\int_{0}^{\epsilon_{c}}\frac{f_{1}\left( \omega\right)
}{\omega^{2}-\epsilon^{2}}\omega d\omega = 0, \label{eq1-mod}
\end{equation}
\begin{equation}
\Lambda\left(  \epsilon\right)  f_{1}\left(  \epsilon\right)
-\pi\Phi f_{2}\left(  \epsilon\right)
-2\Phi\int_{0}^{\epsilon_{c}}\frac{f_{1}\left( \omega\right)
}{\omega^{2}-\epsilon^{2}}\omega d\omega=F_{1}\left(
\epsilon\right),  \label{eq2-mod}
\end{equation}
\begin{equation}
\pi\Phi f_{1}\left(  \epsilon\right)  +\Lambda\left( \epsilon\right)
f_{2}\left(  \epsilon\right)
-2\Phi\epsilon\int_{0}^{\epsilon_{c}}\frac {f_{2}\left(
\omega\right)  }{\omega^{2}-\epsilon^{2}}d\omega-\Phi\int
_{0}^{\epsilon_{c}}\frac{f_{0}\left(  \omega\right)
}{\omega+\epsilon} d\omega = F_{2}\left(  \epsilon\right),
\label{eq3-mod}
\end{equation}
where
\begin{equation}
F_{2}\left(  \epsilon\right) =\frac{1}{2\pi}\int_{0}^{\infty}\left [
\frac{1}{\omega+\epsilon}\frac {\partial N_{0}\left( \omega\right)
}{\partial\omega}-\frac{1}{\omega-\epsilon}\frac {\partial
D_{0}\left( \omega\right) }{\partial\omega}\right ]d\omega.
\label{F_2-mod}
\end{equation}
At $\epsilon>\epsilon_{c}$ we have
\begin{equation}
f_{1}\left(  \epsilon\right) = \frac {1}{2}F_{1}\left(
\epsilon\right)  \approx-\left( 4\epsilon\right)  ^{-1}, \hspace{0.1
cm} F_{2}\left(  \epsilon\right) = 0,
\end{equation}
so $f_{2}\left( \epsilon\right) = 0$ in this range. Yet, no definite
\emph {a priori} information on $f_{0}\left( \epsilon\right) $ at
$\epsilon>\epsilon_{c}$ can be deduced in the considered model.

This set of singular integral equations can be solved using the
well-developed methods \cite{singular} of complex calculus, which is
presented in Appendix. Using Eq. (\ref{f2-f0-eps}) from Appendix, we
get for the function which directly determines the Zitterbewegung
conductivity in Eq. (\ref{sigmaxx1})
\begin{align}
f_{2}\left(  \epsilon\right)  -f_{0}\left(  \epsilon\right)   &
=-\frac{2\left( \pi\Phi\right)  ^{2} }{\left(  2\pi\Phi\right)
^{2}+\Lambda^{2}\left(  \epsilon\right)  } \frac{F_{1}\left(
\epsilon\right)  }{\pi\Phi}+\frac{\Lambda\left( \epsilon\right)
}{\sqrt{\Lambda^{2}\left(  \epsilon\right) +\left(  2\pi
\Phi\right)  ^{2}}}\nonumber\\
&  \times\frac{2}{\pi}\int_{0}^{\epsilon_{c}}\frac{e^{\Theta\left(
\epsilon^{2}\right)  -\Theta\left(  \omega^{2}\right) }F_{1}\left(
\omega\right)  }{\sqrt{\Lambda^{2}\left( \omega\right)  +\left(
2\pi \Phi\right)  ^{2}}}\frac{\omega
d\omega}{\omega^{2}-\epsilon^{2}} -\frac{2\left(  \pi\Phi\right)
^{2}f_{0}\left(  \epsilon\right) }{\left(
2\pi\Phi\right)  ^{2}+\Lambda^{2}\left(  \epsilon\right)  }\nonumber\\
&  -\frac{2\Phi\Lambda\left(  \epsilon\right)
}{\sqrt{\Lambda^{2}\left( \epsilon\right)  +\left( 2\pi\Phi\right)
^{2}}}\int_{0}^{\epsilon_{c}} \frac{e^{\Theta\left(
\epsilon^{2}\right)  -\Theta\left( \omega^{2}\right) }f_{0}\left(
\omega\right) }{\sqrt{\Lambda^{2}\left(  \omega\right)  +\left(
2\pi\Phi\right) ^{2}}}\frac{\omega
d\omega}{\omega^{2}-\epsilon^{2}}. \label{ZBcond-eps}
\end{align}
In turn, $f_{0}\left(  \epsilon\right) $ satisfies Eq.
(\ref{eq-f0-fin}) which is closed Fredholm like integral equation
with a kernel non-singular at $\epsilon = \omega$. As we will see
immediately below, the first term in Eq. (\ref{ZBcond-eps}) is
compatible to a Zitterbewegung one obtained from qualitative
analysis of the Kubo formula for ideal Dirac fermions. \cite{zitter}
The last two terms give rise a novel Zitterbewegung contribution to
$\sigma$, which results in post-leading corrections, $O\left(
1\right)$ at most, for $\Phi\rightarrow 0$ (see Appendix) and so
neglected here. Note that, though local in the energy, the first
term in Eq. (\ref{ZBcond-eps}) is just a result of solving the
singular integral equations system, Eqs. (\ref{eq1-mod}) -
(\ref{eq3-mod}), and not of plain approach when $f_{0}\left(
\epsilon\right) $ and all the singular integral terms of the system
are neglected in advance.

Using the adopted approximation in Eq.(\ref{sigmaxx}) yields for
the Boltzmann conductivity part in the units $e^{2}/h$ the
following \cite{shon}
\begin{equation}
\sigma_{\text{B}}=
-\frac{1}{2\pi\Phi}\int_{0}^{\infty}\frac{\partial N_{0}\left(
\epsilon\right) }{\partial\epsilon}d\epsilon=\frac{N_{0}\left(
0\right) }{2\pi\Phi}=\frac{1}{2\pi\Phi}, \label{sigmaB}
\end{equation}
up to the terms $O(T\mbox{min}(\mu^{-1},\epsilon_{c}^{-1})$. Let
us now integrate the first term in right-hand side of Eq.
(\ref{ZBcond-eps}) over $\epsilon$ using integration by parts.
Thus we obtain the Zitterbewegung contribution to $\sigma$ in a
pseudo-Boltzmann form. Using Eq. (\ref{F}), we have
\begin{align}
& \sigma_{\text{ZB}} =
\sigma_{\text{B}}\int_{0}^{\epsilon_{c}}\frac{\left(
\pi\Phi\right) ^{2}}{\left(  2\pi\Phi\right)
^{2}+\Lambda^{2}\left(  \epsilon\right) }\left[
\frac{1-N_{0}\left(  \epsilon\right)  }{\epsilon}-\frac {\partial
D_{0}\left(  \epsilon\right) }{\partial\epsilon}\right] d\epsilon \nonumber \\
& = -\int_{0}^{\infty}\left[  \epsilon\tau_{\text{ZB}}\left(
\epsilon\right) \frac{\partial N_{0}\left(  \epsilon\right)
}{\partial\epsilon} - \epsilon
\frac{\partial\epsilon\tau_{\text{ZB}}\left( \epsilon\right) }
{\partial\epsilon}\frac{\partial D_{0}\left( \epsilon\right)  }
{\partial\epsilon}\right]  d\epsilon , \label{sigmaZB}
\end{align}
where, by the definition,
\begin{equation}
\tau_{\text{ZB}}\left(  \epsilon\right) =
\frac{\sigma_{\text{B}}}{\epsilon}
\int_{\epsilon}^{\epsilon_{c}}\frac{\left(  \pi\Phi\right)
^{2}}{\left( 2\pi\Phi\right)  ^{2}+\Lambda^{2}\left( \omega\right)
}\frac{d\omega} {\omega} \label{ZBtau}
\end{equation}
is an effective Zitterbewegung relaxation time.
\subsection{Applicability of classical Boltzmann
equation and analysis of the $\mu=0$ case }

From Eq.(\ref{DDelta}) we find at $\epsilon \ll \epsilon_c$
\begin{equation}
\Lambda\left(  \epsilon\right)  \approx2\left(
1+2\Phi\ln\frac{\epsilon }{\epsilon_{c}}\right)  \label{Lambda1}
\end{equation}
i.e. $\Lambda\left(  \epsilon\right)$ zeroes at the energy
\begin{equation}
\epsilon_{\text{K}} = \epsilon_{c}e^{-\frac{1}{2\Phi}} =
\epsilon_{c}e^{-\pi \sigma_{\text{B}}} \label{qKondo}
\end{equation}
which is a striking analog of the Kondo energy scale in a problem of
magnetic impurity in metals. \cite{kondo} Existence of this
exponentially small energy scale in the problem under consideration
was established first in Refs.\onlinecite{D1, D2}. Estimations of
contributions from Eqs. (\ref{sigmaZB}) and Eqs. (\ref{ZBtau}) as
well the neglected Zitterbewegung terms show that at
\begin{equation}
\epsilon _{c}\gg\left\vert \mu\right\vert \gg\max\left(
\epsilon_{\text{K}},T\right) \label{criterion}
\end{equation}
the corrections to the Bornian conductivity (\ref{sigmaB}) are at
most finite in the limit $\Phi \rightarrow 0$ and thus can be
neglected in comparison with $\sigma_{\text{B}}$. This justifies
using the classical Boltzmann equation for graphene, except the
case of extremally small doping.

Formally speaking, application of the theory developed here to the
case of zero doping is doubtful. For example, the self-consistent
Born approximation \cite{shon} gives for this case drastically
different results in comparison with the Born approximation. At
the same time, our approach is formally exact in a sense of
perturbation theory at $\Phi \rightarrow 0$. We will see that,
actually, the classical Born-approximation Boltzmann equation does
not take into account properly all terms of order of $\Phi^{-1}$,
a part of such results from Zitterbewegung.

Let us now perform integration in Eq. (\ref{ZBtau}) assuming
validity of Eq. (\ref{Lambda1}), which is fairly justified at
$\Phi \rightarrow 0$. This yields for the case $\mu=0$
\begin{equation}
\epsilon\tau_{\text{ZB}}\left(  \epsilon\right)
=\frac{\pi\sigma_{B}} {8}\left[
\arctan\frac{1}{\Phi\pi}-\arctan\left
(\frac{2}{\pi}\ln\frac{\epsilon} {\epsilon_{\text{K}}}\right
)\right]. \label{tauZB}
\end{equation}
Substituting Eq. (\ref{tauZB}) into Eq. (\ref{sigmaZB}), we obtain
the integral formula for the conductivity in the undoped graphene
($\mu = 0$)
\begin{align}
\sigma = \sigma_{B}\left\{  1+\frac{\pi}{8}\int_{0}^{\infty}\left[  \frac{\pi}%
{2}-\arctan\left(  \frac{2}{\pi}\ln\frac{2T}{\epsilon_{K}}x\right)
\right] \frac{dx}{\cosh^{2}x}\right\}. \label{sigma-int}
\end{align}
Given $\Phi \ll 1$, this formula has the following asymptotic
behavior with respect to $T$
\begin{equation}
\sigma\approx\sigma_{B}\left\{
\begin{array}
[c]{c}
1+\frac{\pi^{2}}{4}+\frac{\pi^{2}}{16}\left(  \ln\frac{\epsilon_{\text{K}}}%
{T}\right)  ^{-1},\;T\ll\epsilon_{\text{K}}\\
1+\frac{\pi^{2}}{16}\left( \ln\frac{T}{\epsilon_{\text{K}}}\right)
^{-1},\;T\gg\epsilon_{\text{K}}.
\end{array}
\right. \label{asymT}
\end{equation}
It is seen that the Zitterbewegung correction at $T = 0$ has the
same order of magnitude as the Bornian conductivity and numerically
even larger than it by a factor $\pi^2 /4$. The
temperature-dependent corrections are reminiscent to those in the
early theories of the Kondo effect by Abrikosov and Hamann.
\cite{kondo} Thus we obtain the following conductivity ratio
\begin{equation}
\frac{\sigma\left( 0\right)  }{\sigma \left( \infty\right)}= 1 +
\frac{\pi^2}{4}. \label{ratio}
\end{equation}

\section{Conclusions}

We have derived the second-order perturbational GKE for 2D massless
Dirac fermions in graphene scattered by scalar impurity potential.
We considered the GKE solution in the Dirac-delta potential model
with the ultraviolet energy cutoff. Our principal result is the
criterion given by Eq. (\ref{criterion}), which justifies using the
classical Boltzmann equation, except for exponentially narrow
interval of chemical potential and temperature. Our approach clearly
demonstrated that the problem of conductivity at zero doping it
fairly similar to the Kondo problem. In this case, we obtained the
temperature dependent conductivity formula, Eq. (\ref{sigma-int}),
which interpolates well between high-temperature ($T \gg
\epsilon_{\text{K}}$) and low-temperature ($T \ll
\epsilon_{\text{K}}$) ranges and remains finite at $T =
\epsilon_{\text{K}}$. Thus consistent asymptotic solving the
integral equations that result from the derived GKE, we performed in
this paper, proves equivalent to a partial summation of the
perturbation terms $\propto \ln T$. Similar procedure was carried
out for canonical Kondo model using the NSO method in Ref.
\onlinecite{kalaus}.

By the noted analogy with the Kondo problem,\cite{kondo} Eq.
(\ref{sigma-int}) at $T > \epsilon_{\text{K}}$ is asymptotically
correct in the controllable small parameter $\Phi$, while we may not
pretend to describe by it the low-temperature properties, in
particular $\sigma \left (0 \right)$ in detail. Nevertheless for all
$T$, our kinetic equations by construction are more general than the
Boltzmann one (even with scattering rate modified due to chirality
of the current carriers). Therefore, the discrepancy by factor $\sim
3.5$ in the values of $\sigma \left (0 \right)$ obtained, see Eq.
(\ref{ratio}), makes the Boltzmann equation probably not very good
starting point for generalizations, such as the self-consistent Born
approximation.\cite{shon} Would the Kondo analogy goes pretty far,
the observed \cite{r1} $\sigma \left (0 \right) \sim e^2/h$ might be
an evidence that the system enters at $T \ll \epsilon_{\text{K}}$ a
non-perturbational strong effective coupling regime, where the
conductivity attains so called unitary limit,\cite{kondo} rather
than the result of strong bare coupling $\Phi \gg 1$ adopted in Ref.
\onlinecite{shon}.

\section*{Acknowledgements}
This work was supported by the Stichting voor Fundamenteel
Onderzoek der Materie (FOM), the Netherlands, and by the
EuroMagNET.

\section*{Appendix}
Here we present the solution of integral equations
(\ref{eq1-mod})-(\ref{eq3-mod}). Subtracting Eq.(\ref{eq1-mod})
from Eq.(\ref{eq2-mod}) we obtain \textit{purely algebraic} linear
equation
\begin{equation}
\Lambda\left(  \epsilon\right)  f_{1}\left(  \epsilon\right)
-2\pi\Phi f_{2}\left(  \epsilon\right)  -\pi\Phi f_{0}\left(
\epsilon\right) =F_{1}\left(  \epsilon\right)    . \label{eq-allf}
\end{equation}

Let us exclude now $f_{1}\left(  \epsilon\right)  $ expressing it
via $f_{0}\left(  \epsilon\right)  $ and $f_{2}\left(
\epsilon\right) $. To this aim, we use Eq.(\ref{eq1-mod}) in the
form of problem of inverting the Cauchy integral
\begin{equation}
\frac{2}{\pi}\int_{0}^{\epsilon_{c}}\frac{f_{1}\left(
\omega\right)  } {\omega^{2}-\epsilon^{2}}\omega
d\omega=f_{0}\left(  \epsilon\right) +f_{2}\left(  \epsilon\right)
\equiv p_{1}\left( \epsilon\right)  . \label{eqf1-f0f2}
\end{equation}
In what follows we make use of the celebrated Poincare-Bertrand
permutation formula~\cite{singular} for the Cauchy-type integrals
along a contour $C$ in complex plane
\begin{equation}
\int_{C}\left[  \frac{\psi\left(  \epsilon,\tau\right)
}{\tau-\epsilon} \int_{C}\frac{\varphi\left(  \tau,\omega\right)
}{\omega-\tau}d\omega\right] d\tau=\int_{C}\left[
\int_{C}\frac{\psi\left(  \epsilon,\tau\right) \varphi\left(
\tau,\omega\right)  }{\left(  t_{1}-t\right)  \left(
\omega-\tau\right)  }d\tau\right]  d\omega-\pi^{2}\psi\left(
\epsilon ,\epsilon\right)  \varphi\left(  \epsilon,\epsilon\right)
. \label{P-B}
\end{equation}
Putting in Eq. (\ref{eqf1-f0f2}) $\epsilon\rightarrow\tau$,
$\frac{2\omega }{\pi}\frac{f_{1}\left(  \omega\right)
}{\omega+\tau}=\varphi\left( \tau,\omega\right) $, multiplying it
by
\[
\frac{\psi\left(  \epsilon,\tau\right)
}{\tau-\epsilon}=\frac{1}{\left( \tau^{2}-\epsilon^{2}\right)
\sqrt{\epsilon_{c}^{2}-\tau^{2}}}
\]
and integrating over $\tau$ with the use Eq. (\ref{P-B}), we
obtain
\begin{align*}
\int_{0}^{\epsilon_{c}}\frac{p_{1}\left(  \tau\right)
d\tau}{\left(  \tau ^{2}-\epsilon^{2}\right)
\sqrt{\epsilon_{c}^{2}-\tau^{2}}}  &  =-\frac{\pi f_{1}\left(
\epsilon\right)  }{2\epsilon\sqrt{\epsilon_{c}^{2}-\epsilon^{2}}
}+\frac{2}{\pi}\int_{0}^{\epsilon_{c}}\int_{0}^{\epsilon_{c}}\frac{1}{\left(
\tau^{2}-\epsilon^{2}\right)  \left(  \omega^{2}-\tau^{2}\right)
}\frac {d\tau}{\sqrt{\epsilon_{c}^{2}-\tau^{2}}}f_{1}\left(
\omega\right)  \omega
d\omega\\
&  =-\frac{\pi f_{1}\left(  \epsilon\right)
}{2\epsilon\sqrt{\epsilon_{c}
^{2}-\epsilon^{2}}}-\frac{2}{\pi}\int_{0}^{\epsilon_{c}}\frac{I\left(
\omega^{2}\right)  -I\left(  \epsilon^{2}\right)
}{\omega^{2}-\epsilon^{2}
}\frac{d\tau}{\sqrt{\epsilon_{c}^{2}-\tau^{2}}}\frac{f_{1}\left(
\omega\right)  }{\omega^{2}-\epsilon^{2}}\omega d\omega,
\end{align*}
where (the integral below is the principal-value one)
\begin{align*}
I\left(  \epsilon^{2}\right)   &
=\int_{0}^{\epsilon_{c}}\frac{d\tau}{\left(
\tau^{2}-\epsilon^{2}\right)
\sqrt{\epsilon_{c}^{2}-\tau^{2}}}=\frac
{1}{\epsilon_{c}^{2}}\int_{0}^{\frac{\pi}{2}}\frac{d\chi}{\cos^{2}\chi
-\frac{\epsilon^{2}}{\epsilon_{c}^{2}}}=\frac{1}{\epsilon^{2}}\int_{0}
^{\infty}\frac{dt}{\frac{\epsilon_{c}^{2}-\epsilon^{2}}{\epsilon^{2}}-t^{2}}\\
&  =\frac{1}{\epsilon\sqrt{\epsilon_{c}^{2}-\epsilon^{2}}}\int_{0}^{\infty
}\frac{dx}{1-x^{2}}=0,
\end{align*}
from which we deduce that general solution is
\begin{equation}
f_{1}\left(  \epsilon\right)
=-\frac{2\epsilon\sqrt{\epsilon_{c}^{2}
-\epsilon^{2}}}{\pi}\int_{0}^{\epsilon_{c}}\frac{p_{1}\left(
\omega\right) d\omega}{\left(  \omega^{2}-\epsilon^{2}\right)
\sqrt{\epsilon_{c}^{2}
-\omega^{2}}}+\frac{C_{1}}{\epsilon\sqrt{\epsilon_{c}^{2}-\epsilon^{2}}}
\label{f1-gen-sol}
\end{equation}
provided that $p_{1}\left(  \epsilon\right)$ is supposed to be
known and $C_{1}$ is an arbitrary constant. Consider now the
behavior of the above solution at the interval ends. We have
identically
\begin{align*}
f_{1}\left(  \epsilon\right)   &
=-\frac{2\epsilon\sqrt{\epsilon_{c}
^{2}-\epsilon^{2}}}{\pi}\int_{0}^{\frac{\pi}{2}}\frac{p_{1}\left(
\epsilon_{c}\cos\alpha\right)
}{\epsilon_{c}\cos^{2}\alpha-\epsilon^{2}
}d\alpha+\frac{C_{1}}{\epsilon\sqrt{\epsilon_{c}^{2}-\epsilon^{2}}}\\
&
=-\frac{2\epsilon\sqrt{\epsilon_{c}^{2}-\epsilon^{2}}}{\pi}\int_{0}
^{\infty}p_{1}\left(  \frac{\epsilon_{c}}{\sqrt{1+t^{2}}}\right)
\frac
{dt}{\epsilon_{c}^{2}-\epsilon^{2}-\epsilon^{2}t^{2}}+\frac{C_{1}}
{\epsilon\sqrt{\epsilon_{c}^{2}-\epsilon^{2}}}\\
&  =-\frac{2}{\pi}\int_{0}^{\infty}p_{1}\left(
\frac{\epsilon_{c}}
{\sqrt{1+\frac{\epsilon_{c}^{2}-\epsilon^{2}}{\epsilon^{2}}u^{2}}}\right)
\frac{du}{1-u^{2}}+\frac{C_{1}}{\epsilon\sqrt{\epsilon_{c}^{2}-\epsilon^{2}}}.
\end{align*}
Thus, if $p_{1}\left(  0\right)  $ and $p_{1}\left(
\epsilon_{c}\right)  $ are finite, $f_{1}\left(  \epsilon\right) $
at the ends diverges if $C_{1}\neq0$ and is zero if $C_{1}=0$
since $\int_{0}^{\infty}\frac {du}{1-u^{2}}=0$. \ Choosing
$C_{1}=0$, we obtain from Eqs. (\ref{eq-allf}) and
(\ref{f1-gen-sol}) for the function
\begin{equation}
q_{1}\left(  \omega\right)  =\frac{p_{1}\left(  \omega\right)
}{\omega \sqrt{\epsilon_{c}^{2}-\omega^{2}}} \label{fun-q1}
\end{equation}
the following singular integral equation
\begin{align}
\Lambda\left(  \epsilon\right)
\frac{2}{\pi}\int_{0}^{\epsilon_{c}} q_{1}\left(  \omega\right)
\frac{\omega d\omega}{\omega^{2}-\epsilon^{2} }+2\pi\Phi
q_{1}\left(  \epsilon\right)   &  =\frac{\pi\Phi f_{0}\left(
\epsilon\right)  -F_{1}\left( \epsilon\right)
}{\epsilon\sqrt{\epsilon_{c}^{2}-\epsilon^{2}}}\nonumber\\
&  \equiv p_{2}\left(  \epsilon\right)  . \label{eq-q1}
\end{align}
Introducing new variables and functions by
\begin{equation}
\sqrt{x}=\epsilon,\;\sqrt{y}=\omega,\;\widehat{q}_{1}\left(
x\right) =q_{1}\left(  \sqrt{x}\right)  ,\;\widehat{p}_{2}\left(
x\right) =p_{2}\left(  \sqrt{x}\right)  , \label{nvar-q1}
\end{equation}
we convert Eq.(\ref{eq-q1}) to the standard singular integral
equation~\cite{singular}
\begin{equation}
\Lambda\left(  \sqrt{x}\right)
\frac{1}{\pi}\int_{0}^{x_{c}}\frac{\widehat {q}_{1}\left(
y\right)  }{y-x}dy+2\pi\Phi\widehat{q}_{1}\left(  x\right)
=\;\widehat{p}_{2}\left(  x\right)  ,\;0<x<x_{c}=\epsilon_{c}^{2},
\label{eq1-sing}
\end{equation}
assuming $p_{2}\left(  x\right)  $ is known. Following the
standard procedure~\cite{singular}, we define the function of
complex variable
\begin{equation}
Q_{1}\left(  z\right)  =\frac{1}{2\pi
i}\int_{0}^{x_{c}}\frac{\widehat{q} _{1}\left(  y\right)
}{y-z}dy, \label{funX1}
\end{equation}
which is analytic in the plane with the cut along $\left(  0,x_{c}\right)  $
and
\begin{equation}
\lim_{\left\vert z\right\vert \rightarrow\infty}Q_{1}\left(
z\right)  =0. \label{X_1-zinf}
\end{equation}
The relations at $z\rightarrow x\pm0$
\begin{equation}
\widehat{q}_{1}\left(  x\right)  =Q_{1}^{+}\left(  x\right)
-Q_{1}^{-}\left( x\right)  ,\;\frac{1}{\pi
i}\int_{0}^{x_{c}}\frac{\widehat{q}_{1}\left( y\right)
}{y-x}dy=Q_{1}^{+}\left(  x\right)  +Q_{1}^{-}\left(  x\right)
\label{jump1}
\end{equation}
map our equation onto the Riemann-Hilbert boundary value problem
\[
i\Lambda\left(  \sqrt{x}\right)  \left[  Q_{1}^{+}\left(  x\right)
+Q_{1} ^{-}\left(  x\right)  \right]  +2\pi\Phi\left[
Q_{1}^{+}\left(  x\right) -Q_{1}^{-}\left(  x\right)  \right]
=\widehat{p}_{2}\left(  x\right)  ,
\]
or
\begin{equation}
Q_{1}^{+}\left(  x\right)  -G_{1}\left(  x\right)  Q_{1}^{-}\left(
x\right) =\frac{\widehat{p}_{2}\left(  x\right)
}{2\pi\Phi+i\Lambda\left(  \sqrt {x}\right)  }, \label{RHp-Q1}
\end{equation}
where
\begin{equation}
G_{1}\left(  x\right)  =\frac{2\pi\Phi-i\Lambda\left(
\sqrt{x}\right)  } {2\pi\Phi+i\Lambda\left(  \sqrt{x}\right)  }.
\label{funG1}
\end{equation}
Proceeding, we are to solve the homogeneous Riemann-Hilbert
problem of searching a regular analytic function $\Omega_{1}\left(
z\right)  $ satisfying
\begin{equation}
\Omega_{1}^{+}\left(  x\right)  =G_{1}\left(  x\right)
\Omega_{1}^{-}\left( x\right)  , \label{RHp-hom1}
\end{equation}
which is considered below. To obtain the solution of the
inhomogeneous problem following Ref.\onlinecite{singular} we
divide Eq.(\ref{RHp-Q1}) by $\Omega_{1} ^{+}\left(  x\right)  $
and using Eq.(\ref{RHp-hom1}) obtain
\[
\frac{Q_{1}^{+}\left(  x\right)  }{\Omega_{1}^{+}\left(  x\right)
} -\frac{Q_{1}^{-}\left(  x\right)  }{\Omega_{1}^{-}\left(
x\right)  }=\frac {1}{\Omega_{1}^{+}\left(  x\right)
}\frac{\widehat{p}_{2}\left(  x\right) }{2\pi\Phi+i\Lambda\left(
\sqrt{x}\right)  },
\]
from which it immediately follows that
\begin{equation}
Q_{1}\left(  z\right)  =\Omega_{1}\left(  z\right)  \left[
\frac{1}{2\pi i}\int_{0}^{x_{c}}\frac{1}{\Omega_{1}^{+}\left(
y\right)  }\frac{\widehat {p}_{2}\left(  y\right)
}{2\pi\Phi+i\Lambda\left(  \sqrt{y}\right)  }
\frac{dy}{y-z}+P_{1}\left(  z\right)  \right]  , \label{R-H1-sol}
\end{equation}
where $P_{1}\left(  z\right)  $ is an analytic function in the
whole plane except may be points $z=0$ and $z=x_{c}$. The values
of $Q_{1}\left( z\right)  $ on real axis allows one to obtain
$\widehat{q}_{1}\left( x\right)  $ and its Cauchy integral using
first of Eq.(\ref{jump1}) along with Eq. (\ref{R-H1-sol}) as
follows
\begin{align*}
\widehat{q}_{1}\left(  x\right)   &  =\left[  \Omega_{1}^{+}\left(
x\right) -\Omega_{1}^{-}\left(  x\right)  \right]  \left[
\frac{1}{2\pi i}\int _{0}^{x_{c}}\frac{1}{\Omega_{1}^{+}\left(
y\right)  }\frac{\widehat{p} _{2}\left(  y\right)
}{2\pi\Phi+i\Lambda\left(  \sqrt{y}\right)  }\frac
{dy}{y-x}+P_{1}\left(  x\right)  \right] \\
+  &  \frac{1}{2}\left[  1+\frac{\Omega_{1}^{-}\left(  x\right)  }{\Omega
_{1}^{+}\left(  x\right)  }\right]  \frac{\widehat{p}_{2}\left(  x\right)
}{2\pi\Phi+i\Lambda\left(  \sqrt{x}\right)  }\\
&  =-\frac{\Lambda\left(  \sqrt{x}\right)  \Omega_{1}^{+}\left(
x\right) }{2\pi\Phi-i\Lambda\left(  \sqrt{x}\right)  }\left[
\frac{1}{\pi}\int _{0}^{x_{c}}\frac{1}{\Omega_{1}^{+}\left(
y\right)  }\frac{\widehat{p} _{2}\left(  y\right)
}{2\pi\Phi+i\Lambda\left(  \sqrt{y}\right)  }\frac
{dy}{y-x}+2iP_{1}\left(  x\right)  \right] \\
&  +\frac{2\pi\Phi\widehat{p}_{2}\left(  x\right)  }{\left(
2\pi\Phi\right) ^{2}+\Lambda^{2}\left(  \sqrt{x}\right)  }
\end{align*}
and
\begin{align*}
\frac{1}{\pi}\int_{0}^{x_{c}}\frac{\widehat{q}_{1}\left(  y\right)  }{y-x}dy
&  =\left[  \Omega_{1}^{+}\left(  x\right)  +\Omega_{1}^{-}\left(  x\right)
\right]  \left[  \frac{1}{2\pi}\int_{0}^{x_{c}}\frac{1}{\Omega_{1}^{+}\left(
y\right)  }\frac{\widehat{p}_{2}\left(  y\right)  }{2\pi\Phi+i\Lambda\left(
\sqrt{y}\right)  }\frac{dy}{y-x}+iP_{1}\left(  x\right)  \right] \\
&  +\frac{i}{2}\left[  1-\frac{\Omega_{1}^{-}\left(  x\right)  }{\Omega
_{1}^{+}\left(  x\right)  }\right]  \frac{\widehat{p}_{2}\left(  x\right)
}{2\pi\Phi+i\Lambda\left(  \sqrt{x}\right)  }\\
&  =\frac{2\pi\Phi\Omega_{1}^{+}\left(  x\right)
}{2\pi\Phi-i\Lambda\left( \sqrt{x}\right)  }\left[
\frac{1}{\pi}\int_{0}^{x_{c}}\frac{1}{\Omega_{1} ^{+}\left(
y\right)  }\frac{\widehat{p}_{2}\left(  y\right)  }{2\pi
\Phi+i\Lambda\left(  \sqrt{y}\right)
}\frac{dy}{y-x}+2iP_{1}\left(  x\right)
\right] \\
+  &  \frac{\Lambda\left(  \sqrt{x}\right)  \widehat{p}_{2}\left(  x\right)
}{\left(  2\pi\Phi\right)  ^{2}+\Lambda^{2}\left(  \sqrt{x}\right)  }.
\end{align*}

Returning to the relevant homogeneous problem we assume that
$\Omega_{1}\left( z\right)  \neq0,\infty$ at $z\neq0$, $x_{c}$.
Thus we arrive at the inhomogeneous problem for
$\ln\Omega_{1}\left(  z\right)$
\[
\ln\Omega_{1}^{+}\left(  x\right)  -\ln\Omega_{1}^{-}\left(
x\right)  =\ln G_{1}\left(  x\right)
=-2i\arctan\frac{\Lambda\left(  \sqrt{x}\right)  } {2\pi\Phi}.
\]
Note that the end-point conditions are
\begin{equation}
\lim_{x\rightarrow0+}\ln G_{1}\left(  x\right)
=i\pi,\;\lim_{x\rightarrow x_{c}-0}\ln G_{1}\left(  x\right)
=-i\pi, \label{funG1-prop}
\end{equation}
where the first limit holds in general case and the second is the
model property. Consider the following Cauchy
integral~\cite{singular}
\[
U_{1}\left(  z\right)  =\frac{1}{2\pi i}\int_{0}^{x_{c}}\frac{\ln G_{1}\left(
x\right)  }{x-z}dx=-\frac{1}{\pi}\int_{0}^{x_{c}}\arctan\frac{\Lambda\left(
\sqrt{x}\right)  }{2\pi\Phi}\frac{dx}{x-z}.
\]
This function satisfies $\lim_{\left\vert z\right\vert
\rightarrow0} U_{1}\left(  z\right)  =0$. It has two regular nodes
at the points $z=0$ and $z=x_{c}$, which is shown using
integration by parts
\begin{align}
U_{1}\left(  z\right)   &  =-\frac{1}{\pi}\left.  \ln\left(  x-z\right)
\arctan\frac{\Lambda\left(  \sqrt{x}\right)  }{2\pi\Phi}\right\vert
_{0+}^{x_{c}-0}+\frac{1}{\pi}\int_{0}^{x_{c}}\ln\left(  x-z\right)  \frac
{d}{dx}\arctan\frac{\Lambda\left(  \sqrt{x}\right)  }{2\pi\Phi}dx\nonumber\\
&  =-\frac{1}{2}\ln\left[  \left(  -z\right)  \left(
x_{c}-z\right)  \right] +2\Phi\int_{0}^{x_{c}}\frac{\ln\left(
x-z\right)  \Lambda^{\prime}\left( \sqrt{x}\right)
}{\Lambda^{2}\left(  \sqrt{x}\right)  +\left(  2\pi \Phi\right)
^{2}}dx. \label{funU1}
\end{align}
Thus the function $e^{U_{1}\left(  z\right)  }$ can be taken for
$\Omega _{1}\left(  z\right)  $, which satisfies $\lim_{\left\vert
z\right\vert \rightarrow\infty}\Omega_{1}\left(  z\right)  =1$ and
hence it should sustain $\lim_{\left\vert z\right\vert
\rightarrow\infty}P_{1}\left(  z\right)  =0$. For this
$\Omega_{1}\left(  z\right)  $ we have on real axis
\begin{align}
\Omega_{1}^{\pm}\left(  x\right)   &  =e^{U_{1}\left(  x\pm i0\right)
}=e^{-\frac{1}{\pi}\int_{0}^{x_{c}}\arctan\frac{\Lambda\left(  \sqrt
{y}\right)  }{2\pi\Phi}\frac{dy}{y-x\mp i0}}=e^{-\Theta\left(  x\right)  \mp
i\arctan\frac{\Lambda\left(  \sqrt{x}\right)  }{2\pi\Phi}}\nonumber\\
&  =\frac{2\pi\Phi\mp i\Lambda\left(  \sqrt{x}\right)
}{\sqrt{\Lambda ^{2}\left(  \sqrt{x}\right)  +\left(
2\pi\Phi\right)  ^{2}}}e^{\Theta _{1}\left(  x\right)  },
\label{Omega1pm}
\end{align}
where
\begin{equation}
\Theta_{1}\left(  x\right)  =-\frac{1}{\pi}\int_{0}^{x_{c}}\arctan
\frac{\Lambda\left(  \sqrt{y}\right)  }{2\pi\Phi}\frac{dy}{y-x}.
\label{Theta1}
\end{equation}
This yields with $P_{1}=0$
\begin{align}
\widehat{q}_{1}\left(  x\right)   &  =-\frac{\Lambda\left(
\sqrt{x}\right) e^{\Theta_{1}\left(  x\right)
}}{\sqrt{\Lambda^{2}\left(  \sqrt{x}\right) +\left(
2\pi\Phi\right)  ^{2}}}\frac{1}{\pi}\int_{0}^{x_{c}}\frac
{e^{-\Theta_{1}\left(  y\right)  }\widehat{p}_{2}\left(  y\right)
} {\sqrt{\Lambda^{2}\left(  \sqrt{y}\right)  +\left(
2\pi\Phi\right)  ^{2}}
}\frac{dy}{y-x}\nonumber\\
&  +\frac{2\pi\Phi\widehat{p}_{2}\left(  x\right)  }{\left(
2\pi\Phi\right) ^{2}+\Lambda^{2}\left(  \sqrt{x}\right)  }
\label{q1-fin}
\end{align}
and
\begin{align}
\frac{1}{\pi}\int_{0}^{x_{c}}\frac{\widehat{q}_{1}\left(  y\right)
}{y-x}dy &  =\frac{2\pi\Phi e^{\Theta_{1}\left(  x\right)
}}{\sqrt{\Lambda^{2}\left( \sqrt{x}\right)  +\left(
2\pi\Phi\right)  ^{2}}}\frac{1}{\pi}\int_{0}^{x_{c}
}\frac{e^{-\Theta_{1}\left(  y\right)  }\widehat{p}_{2}\left(
y\right) }{\sqrt{\Lambda^{2}\left(  \sqrt{y}\right)  +\left(
2\pi\Phi\right)  ^{2}}
}\frac{dy}{y-x}\nonumber\\
+  &  \frac{\Lambda\left(  \sqrt{x}\right)  \widehat{p}_{2}\left(
x\right) }{\left(  2\pi\Phi\right)  ^{2}+\Lambda^{2}\left(
\sqrt{x}\right)  }. \label{vpq1-fin}
\end{align}
Returning to the energy variables, we obtain
\begin{equation}
p_{1}\left(  \epsilon\right)  =\frac{2\left(  \pi\Phi\right)
^{2}f_{3}\left( \epsilon\right)  }{\left(  2\pi\Phi\right)
^{2}+\Lambda^{2}\left( \epsilon\right)  }-\frac{2\Phi\Lambda\left(
\epsilon\right)  }{\sqrt {\Lambda^{2}\left(  \epsilon\right)
+\left(  2\pi\Phi\right)  ^{2}}}\int
_{0}^{\epsilon_{c}}\frac{e^{\Theta\left(  \epsilon^{2}\right)
-\Theta\left( \omega^{2}\right)  }f_{3}\left(  \omega\right)
}{\sqrt{\Lambda^{2}\left( \omega\right)  +\left(  2\pi\Phi\right)
^{2}}}\frac{\omega d\omega} {\omega^{2}-\epsilon^{2}}
\label{p1-eps}
\end{equation}
and
\begin{equation}
f_{1}\left(  \epsilon\right)  =-\frac{\pi\Phi\Lambda\left(
\epsilon\right) f_{3}\left(  \epsilon\right)  }{\left(
2\pi\Phi\right) ^{2}+\Lambda^{2}\left( \epsilon\right)
}-\frac{\pi\left( 2\Phi\right)  ^{2}}{\sqrt{\Lambda ^{2}\left(
\epsilon\right) +\left(  2\pi\Phi\right)  ^{2}}}\int
_{0}^{\epsilon_{c}}\frac{e^{\Theta\left(  \epsilon^{2}\right)
-\Theta\left( \omega^{2}\right)  }f_{3}\left(  \omega\right)
}{\sqrt{\Lambda^{2}\left( \omega\right)  +\left(  2\pi\Phi\right)
^{2}}}\frac{\omega d\omega} {\omega^{2}-\epsilon^{2}}
\label{f1-eps}
\end{equation}
where
\begin{equation}
f_{3}\left(  \epsilon\right)  =f_{0}\left(  \epsilon\right)
-\frac{F_{1}\left(  \epsilon\right) }{\pi\Phi}, \label{fun-U}
\end{equation}
and
\begin{align}
\Theta\left(  \epsilon^{2}\right)   &  =\Theta_{1}\left(
\epsilon^{2}\right) +\ln\sqrt{\epsilon^{2}\left(
\epsilon_{c}^{2}-\omega^{2}\right)  }=2\Phi
\int_{0}^{\epsilon_{c}}\frac{\Lambda^{\prime}\left(
\sqrt{x}\right) }{\Lambda^{2}\left(  \sqrt{x}\right)  +\left(
2\pi\Phi\right)  ^{2}}
\ln\left\vert x-\epsilon^{2}\right\vert dx\nonumber\\
&
=\Phi^{2}\epsilon_{c}^{2}\int_{0}^{\epsilon_{c}}\frac{\ln\left\vert
x-\epsilon^{2}\right\vert }{\left(
1+\Phi\ln\frac{x}{\epsilon_{c}^{2} -x}\right)  ^{2}+\left(
\pi\Phi\right)  ^{2}}\frac{dx}{x\left(  \epsilon
_{c}^{2}-x\right)  }\nonumber\\
&  =\int_{-\infty}^{\infty}\ln\left\vert
\frac{e^{u}}{e^{u}+1}-\frac
{\epsilon^{2}}{\epsilon_{c}^{2}}\right\vert \frac{du}{\left(  \Phi
^{-1}+u\right)  ^{2}+\pi^{2}} , \label{fun-Theta}
\end{align}
the constant $\Phi^{2}\ln\epsilon_{c}^{2}$ being omited since only
difference $\Theta\left(  \epsilon^{2}\right)  -\Theta\left(
\omega^{2}\right)  $ enters all formulas.

>From Eq. (\ref{p1-eps}) and the definition of $p_{1}\left(
\epsilon\right)  $ we obtain the connection between two functions
of interest
\begin{align}
f_{2}\left(  \epsilon\right)   &  =-\frac{F_{1}\left(
\epsilon\right)  } {\pi\Phi}-\frac{2\left(  \pi\Phi\right)
^{2}+\Lambda^{2}\left( \epsilon\right)  }{\left(  2\pi\Phi\right)
^{2}+\Lambda^{2}\left(
\epsilon\right)  }f_{3}\left(  \epsilon\right) \nonumber\\
&  -\frac{2\Phi\Lambda\left(  \epsilon\right)
}{\sqrt{\Lambda^{2}\left( \epsilon\right)  +\left( 2\pi\Phi\right)
^{2}}}\int_{0}^{\epsilon_{c}} \frac{e^{\Theta\left(
\epsilon^{2}\right)  -\Theta\left( \omega^{2}\right) }f_{3}\left(
\omega\right) }{\sqrt{\Lambda^{2}\left(  \omega\right)  +\left(
2\pi\Phi\right) ^{2}}}\frac{\omega
d\omega}{\omega^{2}-\epsilon^{2}}. \label{f2-f0-eps}
\end{align}
Using Eqs.(\ref{f1-eps}),(\ref{f2-f0-eps}) and (\ref{fun-U}) along
with Eq.(\ref{eq3-mod}) and the Poincare-Bertrand formula, see
Eq.(\ref{P-B}), we obtain closed integral equation for the function
$f_{0}\left(  \epsilon\right)$. The equation reads
\begin{equation}
\Lambda\left(  \epsilon\right)  f_{0}\left(  \epsilon\right)
-\Phi\int _{0}^{\epsilon_{c}}\left[  Q\left(
\epsilon,\omega\right) -\frac {1}{\omega+\epsilon}\right]
f_{0}\left(  \omega\right) d\omega=F_{3}\left( \epsilon\right)  ,
\label{eq-f0-fin}
\end{equation}
where the kernel and inhomogeneity term are given by

\begin{align}
Q\left(  \epsilon,\omega\right)   &
=\frac{1}{\omega^{2}-\epsilon^{2} }\left\{  \frac{\omega\left[
\left(  2\pi\Phi\right)  ^{2}+2\Lambda ^{2}\left(  \epsilon\right)
\right]  e^{\Theta\left(  \epsilon^{2}\right) -\Theta\left(
\omega^{2}\right)  }}{\sqrt{\Lambda^{2}\left(  \epsilon\right)
+\left(  2\pi\Phi\right)  ^{2}}\sqrt{\Lambda^{2}\left(
\omega\right)
+\left(  2\pi\Phi\right)  ^{2}}}\right. \nonumber\\
&  -\left.  \frac{\epsilon\left[  \left(  2\pi\Phi\right)
^{2}+2\Lambda ^{2}\left(  \omega\right)  \right]  }{\left(
2\pi\Phi\right)  ^{2} +\Lambda^{2}\left(  \omega\right)  }\right\}
+\Phi K\left( \epsilon,\omega\right)  \label{kernel-Q1}
\end{align}
and
\begin{equation}
F_{3}\left(  \epsilon\right)  =-F_{2}\left(  \epsilon\right)
-\frac{1}{\pi}\int_{0}^{\epsilon_{c}}\left[ \frac{2\epsilon}
{\omega^{2}-\epsilon^{2}}+Q\left( \epsilon,\omega\right) \right]
F_{1}\left(  \omega\right) d\omega, \label{F_3}
\end{equation}
respectively, while
\begin{equation}
K\left(  \epsilon,\omega\right)
=\frac{4\epsilon\omega}{\sqrt{\Lambda ^{2}\left(  \omega\right)
+\left(  2\pi\Phi\right)  ^{2}}}\int_{0}
^{\epsilon_{c}}\frac{e^{\Theta\left(  \tau^{2}\right)
-\Theta\left( \omega^{2}\right)  }\Lambda\left(  \tau\right)
d\tau}{\left(  \tau ^{2}-\epsilon^{2}\right)  \left(
\omega^{2}-\tau^{2}\right)  \sqrt {\Lambda^{2}\left(  \tau\right)
+\left(  2\pi\Phi\right)  ^{2}}} \label{kernel-K1}
\end{equation}
is another kernel. Note that, like $K\left( \epsilon,\omega\right)
$, the kernel $Q\left( \epsilon,\omega\right)$ is non-singular at
$\omega=\epsilon\neq0$. Also we have
\begin{equation}
\lim_{\Phi\rightarrow0}K\left(  \epsilon,\omega\right) =2\epsilon
\omega\int_{0}^{\epsilon_{c}}\frac{d\tau}{\left( \tau^{2}-\epsilon
^{2}\right)  \left(  \omega^{2}-\tau^{2}\right)
}=\frac{\omega\ln\left(
\frac{\epsilon_{c}+\epsilon}{\epsilon_{c}-\epsilon}\right)
-\epsilon \ln\left(
\frac{\epsilon_{c}+\omega}{\epsilon_{c}-\omega}\right)  }
{\epsilon^{2}-\omega^{2}} \label{K1-asym}
\end{equation}
and
\begin{equation}
\lim_{\Phi\rightarrow0}Q\left(  \epsilon,\omega\right) =2\frac
{\omega-\epsilon}{\omega^{2}-\epsilon^{2}}=\frac{2}{\epsilon+\omega}
=Q^{\left(  0\right)  }\left(  \epsilon,\omega\right)  .
\label{Q1-asym}
\end{equation}
Further, the leading terms in $\Phi$ at $\Phi\ll1$ of the
inhomogeneity term is
\begin{equation}
\lim_{\Phi\rightarrow0}F_{3}\left(  \epsilon\right)  =-F_{2}\left(
\epsilon\right) +\frac{2}{\pi}\int_{0}^{\epsilon_{c} }\frac{
F_{1}\left(  \omega\right) \omega }{\omega^{2}-\epsilon^{2}
}d\omega =F_{3}^{\left(  0\right) }\left( \epsilon\right)  .
\label{F_3-asym}
\end{equation}
It is clearly seen from the above equations that $f_{0}\left
(\epsilon\right ) = O\left (1\right )$ at $\Phi\rightarrow 0$,
which results in the conductivity formula obtained in the main
text.

\end{document}